# Added Value of Intraoperative Data for Predicting Postoperative Complications: Development and Validation of a *MySurgeryRisk* Extension

Shounak Datta, PhD[a, g*]
Tyler J. Loftus, MD[b, g*]
Matthew M. Ruppert, BS[a, g]
Chris Giordano, MD[c]
Lasith Adhikari, PhD[a, g]
Ying-Chih Peng, BS[d, g]
Yuanfang Ren, PhD[a. g]
Benjamin Shickel, MS[e, g]
Zheng Feng, MS[f, g]
Gloria Lipori, MBA[h]
Gilbert R. Upchurch Jr., MD[b]
Xiaolin Li, PhD[f, g]
Parisa Rashidi, PhD[e, g]
Tezcan Ozrazgat-Baslanti, PhD[a, g*]
Azra Bihorac, MD, MS[a, g*]

* These authors have contributed equally
[a] Department of Medicine, University of Florida, Gainesville, FL USA
[b] Department of Surgery, University of Florida, Gainesville, FL USA
[c] Department of Anesthesiology, University of Florida, Gainesville, FL USA
[d] Department of Industrial and Systems Engineering, University of Florida, Gainesville, FL USA
[e] Department of Biomedical Engineering, University of Florida, Gainesville, FL USA
[f] Department of Electrical and Computer Engineering, University of Florida, Gainesville, FL USA
[g] Precision and Intelligent Systems in Medicine (Prisma[P]), University of Florida, Gainesville, FL USA
[h] University of Florida Health, Gainesville, FL USA

**Corresponding author:** Azra Bihorac MD MS, Department of Medicine, Precision and Intelligent Systems in Medicine (Prisma[P]), Division of Nephrology, Hypertension, and Renal Transplantation, PO Box 100224, Gainesville, FL 32610-0224. Telephone: (352) 294-8580; Fax: (352) 392-5465; Email: abihorac@ufl.edu

**Conflicts of Interest Disclosures and Source of Funding:** A.B., T.O.B., and M.R. were supported by R01 GM110240 from the National Institute of General Medical Sciences. A.B. and T.O.B. were supported by Sepsis and Critical Illness Research Center Award P50 GM-111152 from the National Institute of General Medical Sciences. T.O.B. received a grant that was supported by the National Center for Advancing Translational Sciences of the National Institutes of Health under Award Number UL1TR001427 and received grant support from Gatorade Trust (127900), University of Florida. T.J.L. was supported by a post-graduate training grant (T32 GM-008721) in burns, trauma, and perioperative injury from the National Institute of General Medical Sciences. M.R. received support from the University of Florida Davis Foundation. This work was supported in part by the NIH/NCATS Clinical and Translational Sciences Award to the University of Florida UL1 TR000064.

## ABSTRACT

**Objective:** To test the hypothesis that accuracy, discrimination, and precision in predicting postoperative complications improve when using both preoperative and intraoperative data input features versus preoperative data alone.

**Background**: Models that predict postoperative complications often ignore important intraoperative events and physiological changes. Incorporation of intraoperative physiological data may improve model performance.

**Methods**: This retrospective cohort analysis included 43,943 adults undergoing 52,529 inpatient surgeries at a single institution during a five-year period. Random forest machine learning models in the validated *MySurgeryRisk* platform made patient-level predictions for three postoperative complications and mortality during hospital admission using electronic health record data and patient neighborhood characteristics. For each outcome, one model trained with preoperative data alone and one model trained with both preoperative and intraoperative data. Models were compared by accuracy, discrimination (expressed as AUROC: area under the receiver operating characteristic curve), precision (expressed as AUPRC: area under the precision-recall curve), and reclassification indices (NRI).

**Results**: Machine learning models incorporating both preoperative and intraoperative data had greater accuracy, discrimination, and precision than models using preoperative data alone for predicting all three postoperative complications (intensive care unit length of stay >48 hours, mechanical ventilation >48 hours, and neurological complications including delirium) and in-hospital mortality (accuracy: 88% vs. 77%, AUROC: 0.93 vs. 0.87, AUPRC: 0.21 vs. 0.15). Overall reclassification improvement was 2.9-10.0% for complications and 11.2% for in-hospital mortality.

**Conclusions**: Incorporating both preoperative and intraoperative data significantly increased accuracy, discrimination, and precision for machine learning models predicting postoperative complications.

## Introduction

Predicting postoperative complications in the preoperative setting informs the surgeon's decision to offer an operation and the patient's decision to undergo surgery. It also guides prehabilitation and other risk-reduction strategies, plans for postoperative resource use, and expectations for short- and long-term prognosis. Online risk calculators, mobile device applications, and automated predictive analytic platforms accomplish these goals[1-4]. However, these models often ignore intraoperative data, thus missing potentially important opportunities to generate updated predictions that inform decisions regarding postoperative triage, surveillance for complications, and targeted preventative measures.

Although it seems advantageous to use intraoperative data in predicting postoperative complications, this advantage remains theoretical until establishing that predictive accuracy, discrimination, and precision improve by incorporating intraoperative data, and that better predictions translate to better decisions and outcomes. This study addresses the former objective. Added value of intraoperative data has been demonstrated for predicting postoperative AKI[5]. This study assessed added value of intraoperative data for predicting three postoperative complications and mortality by developing and validating a *MySurgeryRisk* extension that incorporates vital signs and mechanical ventilator data collected during surgery. The original *MySurgeryRisk* platform uses electronic health record (EHR) data and patient neighborhood characteristics to predict postoperative complications and mortality, but ignores intraoperative data[4]. We hypothesized that accuracy, discrimination, and precision in predicting postoperative complications and mortality would improve when using both preoperative and intraoperative physiological time-series input data versus preoperative data alone.

## Materials and Methods

We created a single-center longitudinal cohort of surgical patients with data from preoperative, intraoperative, and postoperative phases of care. We used random forest machine learning models to predict three major postoperative complications and death during admission, comparing models using preoperative data (i.e. EHR and patient neighborhood characteristics) alone versus models using the same preoperative data plus intraoperative physiological time-series data. The University of Florida Institutional Review Board and Privacy Office approved this study with waiver of informed consent (IRB #201600223).

### Data Source

The University of Florida Integrated Data Repository was used as an honest broker to assemble a single center longitudinal perioperative cohort for all patients admitted to the University of Florida Health for longer than 24 hours following any type of operative procedure between June 1st, 2014 through March 1st, 2019 by integrating electronic health records with other clinical, administrative, and public databases as previously described [4]. The resulting dataset included detailed information on patient demographics, diagnoses, procedures, outcomes, comprehensive hospital charges, hospital characteristics, insurance status, laboratory, pharmacy, and blood bank data as well as detailed intraoperative physiologic and monitoring data for the cohort.

### Participants

We identified all patients with age 18 years or greater and excluded patients who died during surgery and had incomplete records. If patients underwent multiple surgeries during one admission, only the first surgery was used in our analysis. The final cohort consisted of 43,943 patients undergoing 52,529 surgeries. **Supplemental Digital Content 1** illustrates derivation of the study population. **Supplemental Digital Content 2** illustrates cohort use and purpose.

**Outcomes**

We modeled risk for developing three major postoperative complications and mortality occurring after the index surgery and before hospital discharge. Complications included intensive care unit (ICU) admission and mechanical ventilation (MV) for greater than 48 hours, and neurological complications including delirium.

**Predictor Features**

The risk assessment used 367 demographic, socioeconomic, comorbidity, medication, laboratory value, operative, and physiological variables from preoperative and intraoperative phases of care. Of these 367 variables, 134 were used for preoperative only model and 233 intraoperative features were incorporated for developing postoperative models. We derived preoperative comorbidities from International Classification of Diseases (ICD) codes to calculate Charlson comorbidity indices.[6] We modeled primary procedure type on ICD-9-CM codes with a forest structure in which nodes represented groups of procedures, roots presented the most general groups of procedures, and leaf nodes represented specific procedures. Medications were derived from RxNorm codes grouped into drug classes as previously described.[4] We converted intraoperative time series data into statistical features such as minimum, maximum, mean, and short- and long-term variability[7]. Intraoperative data input features included heart rate, systolic blood pressure, diastolic blood pressure, body temperature, respiratory rate, minimum alveolar concentration (MAC), positive end-expiratory pressure (PEEP), peak inspiratory pressure (PIP), fraction of inspired oxygen (FiO2), blood oxygen saturation (SpO2), and end-tidal carbon dioxide (EtCO2). We also included surgical variables (e.g., nighttime surgery, surgery duration, operative blood loss, and urine output) during surgery. **Supplemental Digital Content 3** lists all input features and their statistical characteristics.

**Sample Size**

Models were trained on a development cohort of 40,560 surgeries. All results were reported from a validation cohort of 11,969 surgeries. We performed five-fold cross-validation using random partitions to generate five disjoint folds, allocating one fold for validation and the other four for training. Using a validation cohort of 11,969 surgeries, the overall sample size allows for a maximum width of the 95% confidence interval for area under the receiver operating characteristic curve (AUROC) to be between 0.02 to 0.04 for postoperative complications with prevalence ranging between 5% and 30% for AUROC of 0.80 or higher. The sample size allows for a maximum width of 0.07 for hospital mortality given a 2% prevalence.

**Predictive Analytic Workflow**

The proposed *MySurgeryRisk* Postop algorithm is conceptualized as a dynamic model that readjusts the preoperative risks using physiological time series and other data collected during surgery. The resulting adjusted postoperative risk is assessed immediately at the end of surgery. This flow simulates the clinical task faced by physicians involved in perioperative care where patients' preoperative information is subsequently enriched by the influx of new data from the operating room. The final output produces *MySurgeryRisk* Postop, a personalized risk panel for AKI after surgery with both preoperative and immediate postoperative risk assessments. The algorithm consists of two main layers, preoperative and intraoperative, each containing two cores, data transformer and data analytics.[4] Briefly, the *MySurgeryRisk* platform uses a *Data Transformer* to integrate data from multiple sources, including the EHR with zip code links to US Census data for patient neighborhood characteristics and distance from the hospital, and optimizes the data for analysis through preprocessing, feature transformation, and feature selection techniques. In the preprocessing step, we replaced missing nominal variables with a distinct ''missing'' category and replaced missing continuous variables with the median value for that variable. **Supplemental Digital Content 5** lists allowable ranges for continuous variables, determined by clinical expertise. In the feature transformation step, we reduced data dimensionality and overfitting by transforming categorical variables with more than two levels (e.g., zip code, surgeon) to numeric values by calculating the conditional predicted probability that a certain categorical variable would be associated with a certain complication. This was represented by the log of the ratio of prevalence of that variable among surgeries with a complication to prevalence of that variable among surgeries without a complication (i.e., surgeon x operated on 6% of all cases with wound complications and 8% of all cases without wound complications = log (0.06/0.0.8)). We used a forest of trees method with supervised feature selection to group similar surgeries by ICD procedure codes with anatomic prefixes[8].

In the *data analytics core*, the *MySurgeryRisk* Postop algorithm was trained to calculate patient-level immediate postoperative risk probabilities for selected complications using all available preoperative and intraoperative data with random forest classifiers[9]. In the first stage, the algorithm was trained to calculate preoperative risk probabilities using preoperative data only. Feature selection and other hyper parameters in the scikit-learn random forest classifier (i.e., number of trees, maximum features for the best split, minimum number of samples required per leaf node) were tuned simultaneously using a grid search technique with 5-fold cross validation[10]. To help the classifier address the unbalanced class distribution in our cohorts, the 'balanced' option was selected for 'class weight' parameter. We designed the classifiers to pursue the highest AUROC possible. We used the models to produce risk scores for aforementioned complications, combined them with prediction results of postoperative Acute Kidney Injury model[5] and Sepsis model[11], and used these probabilities as predictive features for predicting in-hospital mortality using the same random forest approach. **Figure 1** illustrates our method for building the random forest machine learning models and model analytic flow.

**Model Validation**

Results are reported from application of the trained model on the test cohort, with 10,637 unique patients undergoing 11,969 surgeries from March 1st, 2018 through March 1st, 2019 time period. Using the prediction results obtained from the 1000 bootstrap cohorts, nonparametric confidence intervals for each of the performance metrics were calculated.

**Model Performance**

We assessed each model's discrimination using AUROC. For each complication, we calculated Youden's index threshhold to identify the point on the receiver operating characteristic curve with the highest combination of sensitivity and specificity, using this point as the cut-off value for low versus high risk[12]. We used these cut-off values to determine the fraction of correct classifications as well as sensitivity, specificity, positive predictive value, and negative predictive value for each model. When rare events are being predicted, a model can have high accuracy by favoring negative predictions in a predominantly negative dataset[13]. False negative predictions of complications are particularly harmful because patients and their caregivers may consent to an operation under the pretense of an overly optimistic postoperative prognosis, and providers may miss opportunities for preoperative mitigation of risk factors through prehabilitation and other optimization strategies. Therefore, model performance was also evaluated by calculating area under the precision-recall curve (AUPRC), which is well-suited for evaluating rare event predictive performance.[14] To assess the statistical significance of AUROC, AUPRC, and accuracy differences between models, we performed Wilcoxon's Sign-Ranked test[15]. We used bootstrap sampling and non-parametric methods to obtain 95% confidence intervals for all performance metrics. The Net Reclassification Improvement (NRI) index was used to quantify how well the postoperative model reclassifies patients compared to the preoperative model[16].

## Results

### Participant Baseline Characteristics and Outcomes

**Table 1** lists subject characteristics of primary interest. **Supplemental Digital Content 4** lists all additional subject characteristics used to build the models. Approximately 49% of the population was female. Average age was 57 years. The incidence of complications was as follows: 26% for prolonged ICU stay, 6% for mechanical ventilation for >48 hours, 16% for neurological complications, and 2% for in-hospital mortality. The distribution of outcomes did not significantly differ between training and testing cohorts, as listed in **Table 1.**

### Model Performance

Compared with the model using preoperative data alone, the extended model using both preoperative and intraoperative data had higher accuracy, AUROC, and AUPRC for all complications and mortality predictions, as described below and in **Table 2**. Furthermore, model performance comparisons for AUROC and AUPRC are provided visually in **Figure 2**. The net reclassification index along with event, non-event, and overall classification improvement for each outcome are listed in **Table 3**.

**Prolonged ICU Stay**

The extended model achieved greater accuracy (0.83 vs. 0.77, $p<0.001$), discrimination (AUROC 0.88 vs. 0.87, $p<0.001$), and precision (AUPRC 0.80 vs. 0.72, $p<0.001$) in predicting ICU stay > 48 hours with greater specificity and positive predictive value at the cost of lower sensitivity (75% vs. 82%, $p<0.001$) than the model using preoperative data alone (**Table 2**). The extended model misclassified 7.9% of all cases that featured prolonged ICU stays, and correctly reclassified 12.6% of all cases that did not (**Figure 3**). Overall, there was a 6.8% reclassification improvement by the extended model.

**Prolonged Mechanical Ventilation**

The extended model achieved greater accuracy (0.92 vs. 0.82, $p<0.001$), discrimination (AUROC 0.96 vs. 0.89, $p<0.001$), and precision (AUPRC 0.71 vs. 0.45, $p<0.001$) in predicting mechanical ventilation > 48 hours with greater sensitivity, specificity, and positive predictive value, and similar negative predictive value compared with the model using preoperative data alone (**Table 2**). The extended model correctly reclassified 11.0% of all cases that featured prolonged mechanical ventilation and 9.9% of all cases that did not (**Figure 4**). Overall reclassification improvement was 10.0%.

**Neurological Complications and Delirium**

The extended model achieved greater accuracy (0.81 vs. 0.78, $p<0.001$), discrimination (AUROC 0.89 vs. 0.86, $p<0.001$), and precision (AUPRC 0.69 vs. 0.64, $p<0.001$) in predicting postoperative neurological complications and delirium with greater specificity, positive predictive value, and negative predictive value than the model limited to preoperative data alone (**Table 2**). The extended model correctly reclassified 2.1% of all cases that featured postoperative neurological complications and delirium and 3.1% of all cases that did not (**Figure 5**). Overall reclassification improvement was 2.9%.

### In-hospital Mortality

The extended model achieved greater accuracy (0.88 vs. 0.77, p<0.001), discrimination (AUROC 0.93 vs. 0.87, p<0.001), and precision (AUPRC 0.21 vs. 0.15, p<0.001) in predicting postoperative in-hospital mortality with greater specificity and positive predictive value, and similar sensitivity and negative predictive value compared with preoperative data alone (**Table 3**). The extended model correctly reclassified 2.2% of all cases of postoperative in-hospital mortality and 11.5% of all cases in which the patient survived to hospital discharge (**Figure 6**). Overall reclassification improvement was 11.2%

### Time-consumption in Model Training

For one point of grid search (e.g., one value of estimator number, minimum sample leaf number, best k value, and maximum allowable feature number) with 5-fold cross validation, the typical time for model training with both preoperative and intraoperative data was 550 - 690 seconds. Using preoperative data alone, training time was 395-460 seconds.

## Discussion

We found that incorporating intraoperative physiological data added value to a machine learning model predicting postoperative complications by improving accuracy, discrimination, and precision relative to a model using preoperative data alone. This was true for all three postoperative complications tested as well as in-hospital mortality. There were no cases in which accuracy, discrimination, and precision did not improve by incorporating intraoperative data. The only negative consequences occurred when predicting prolonged ICU stay; the extended models had lower sensitivity than the model using preoperative data alone. In this case, it appears that the model using preoperative data alone had unusually low thresholds for classifying patients as high risk. The extended models raised this threshold, correctly classifying a greater proportion of patients and achieving greater accuracy, discrimination, and precision, at the cost of lower sensitivity.

Online risk calculators like the National Surgical Quality Improvement Program (NSQIP) Surgical Risk Calculator can reduce variability and increase the likelihood that patients will engage in prehabilitation, but they have time-consuming manual data acquisition and entry requirements, which hinders their clinical adoption[17-22]. Emerging technologies can circumvent this problem. The *MySurgeryRisk* platform autonomously draws data from multiple input sources and uses machine learning techniques to predict postoperative complications and mortality. However, easily and readily available predictions are only useful if they are accurate and precise enough to augment clinical decision-making. In a prospective study of the original *MySurgeryRisk* platform, the algorithm predicted postoperative complications with greater accuracy than physicians, but left room for improvement[23]. The present study demonstrates that incorporation of intraoperative physiological time-series data improves predictive accuracy, discrimination, and precision, presumably by representing important intraoperative events and physiological changes that influence postoperative clinical trajectories and complications. Recently, Dziadzko et al.[24] used a random forest model to predict death or mechanical ventilation for greater than 48 hours using EHR data from patients admitted to academic hospitals, achieving excellent discrimination (AUROC 0.90), similar to *MySurgeryRisk* discrimination for mechanical ventilation for greater than 48 hours (AUROC 0.96) using both preoperative and intraoperative data. Therefore, this extension of the *MySurgeryRisk* platform takes another step toward clinical utility, maintaining autonomous function while improving accuracy, discrimination, and precision.

Despite advances in ease of use and performance, predictive analytic platforms face another barrier to clinical adoption; predictions are not decisions. When predicted risk for postoperative AKI is very low or very high, it is relatively clear whether the patient would benefit from renal-protection bundles. Similarly, when predicted risk for cardiovascular complications is very low or very high, it is relatively clear whether the patient would benefit from continuous cardiac monitoring. However, a substantial number of patients are at intermediate risk for these complications. In the present study, we dichotomized outcome predictions into low- and high-risk categories to facilitate analysis of model performance, but risk for complications exists on a continuum. The *MySurgeryRisk* platform makes predictions along a continuum (i.e., range from 0%-100% chance of a complication), but this method also fails to augment clinical decisions for intermediate-risk scenarios. Average risk across a population usually defines intermediate risk. Therefore, this challenge affects most patients.

Advances in machine learning technologies may rise to meet this challenge. Predictive analytics indirectly inform discrete choices facing clinicians; reinforcement learning models can provide instructive feedback, identifying specific actions yielding the highest probability of achieving a defined goal. For example, a reinforcement learning model could be trained to achieve hospital discharge with baseline renal and cardiovascular function, without major adverse kidney or cardiac events, making recommendations for or against renal protection bundles and continuous cardiac monitoring according to these goals. Similar models have been used to recommend vasopressor doses and intravenous fluid resuscitation volumes for septic patients, demonstrating efficacy relative to clinician decision-making in large retrospective datasets[25]. However, to our knowledge, these models have not been tested clinically or applied to surgical decision-making scenarios. Therefore, the potential benefits of reinforcement learning to augment surgical decision-making learning remain theoretical.

This study used data from a single institution, limiting the generalizability of these findings. As previously discussed, true risk for complications is not dichotomous, but we dichotomized risk in this study to facilitate model performance evaluation and comparison. We used administrative codes to identify complications, so coding errors could have influenced results. The *MySurgeryRisk* algorithm learned predictive features from raw data, and so it may have used features that are not classic risk factors. This approach has the potential advantage of discovering and incorporating unknown or underused risk factors, and the disadvantage that the existence and identity of these risk factors remain unknown.

## **Conclusions**

Incorporation of both preoperative and intraoperative data significantly increased the accuracy, discrimination, and precision of machine learning models predicting three postoperative complications and in-hospital mortality. The added value of intraoperative data was universal with the exception of a 7% decrease in sensitivity for predicting prolonged ICU stay. These predictions have the theoretical benefit of supporting decisions regarding postoperative triage, surveillance for complications, and targeted preventative measures. Future research should seek to augment decision-making for intermediate-risk patients, who represent most postoperative patients and pose the greatest decision-making challenges.

## Figure legends

**Figure 1: Conceptual diagram of the Postoperative *MySurgeryRisk* Complication prediction model.** This diagram shows the aggregation of the data transformer, data engineering, and data analytics modules in the preoperative and intraoperative layers. The two layers are integrated by obtaining the full perioperative dataset by merging all the clean features from both layers (orange arrow)

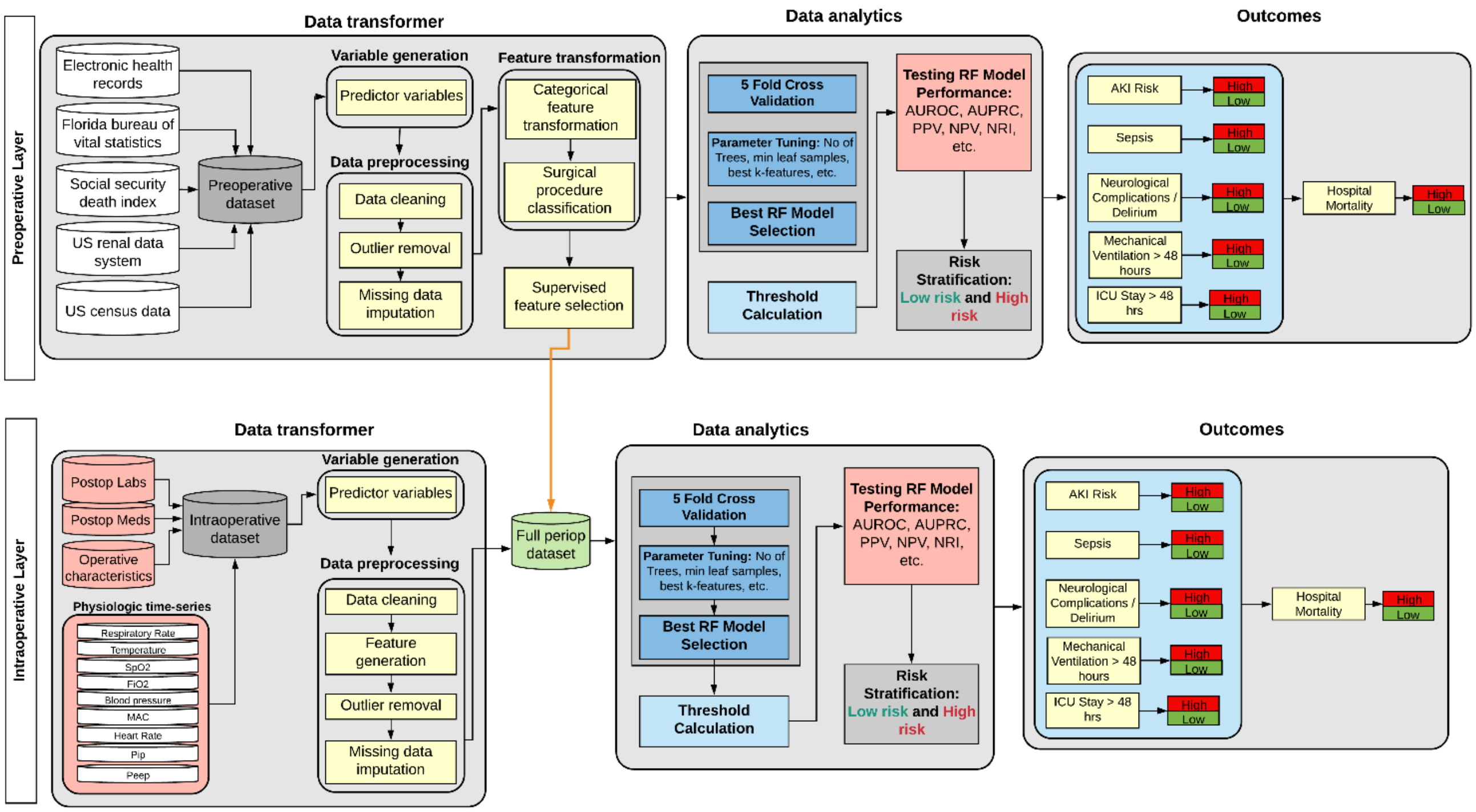

**Figure 2: Models using both preoperative and intraoperative data had greater discrimination and precision than models using preoperative data alone in predicting three postoperative complications and mortality.** Comparison of (A)AUROC and (B) AUPRC for all predicted outcomes. AUROC: area under the receiver operating characteristic curve, AUPRC: area under the precision-recall curve, ICU: intensive care unit, MV: mechanical ventilation.

**(A)**

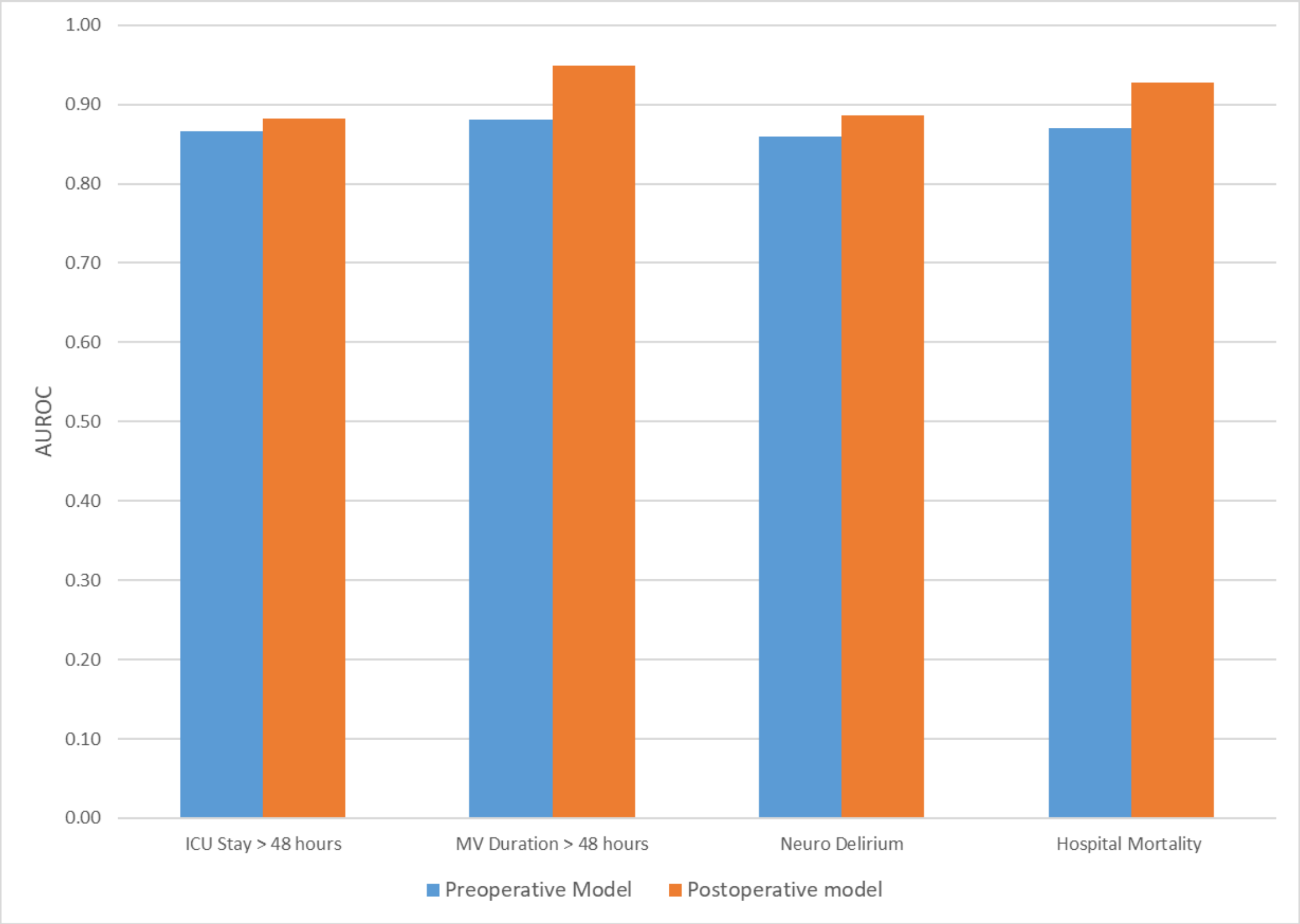

**(B)**

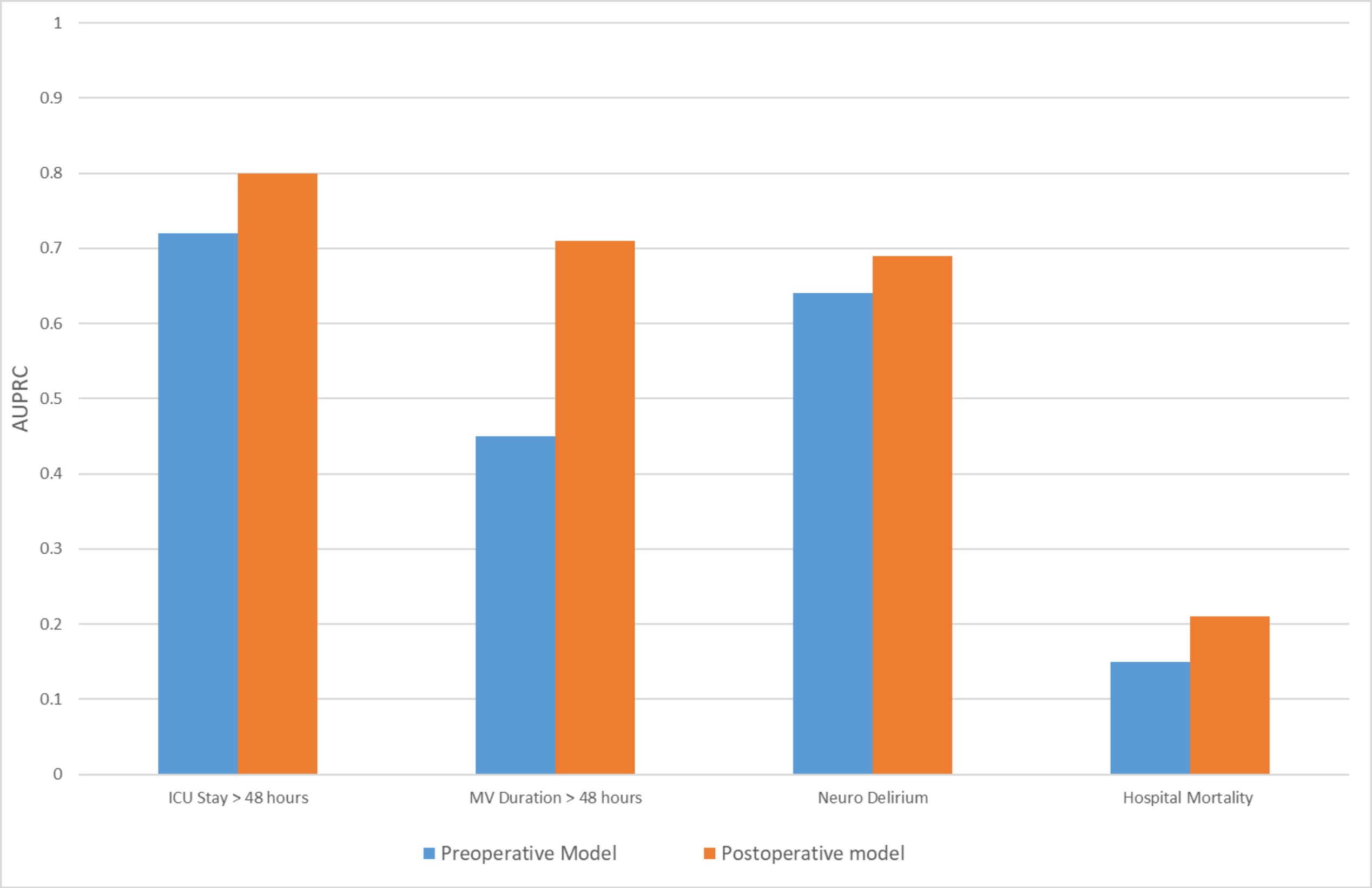

**Figure 3: A model using both preoperative and intraoperative data outperformed a model using preoperative data alone in predicting postoperative ICU stay >48 hours.** A: The postoperative model had greater area under the receiver operating characteristic curve (0.88 vs. 0.87). B: The postoperative model had greater area under the precision-recall curve (0.80 vs. 0.72). The postoperative model reclassified positive cases of prolonged ICU stays (C) and negative cases (D). The red dots are patients at high-risk for prolonged ICU stay according to the postoperative model, whereas the green dots are patients at low-risk. (C & D) The proposed postoperative model effectively reclassified 6.8% of all cases.

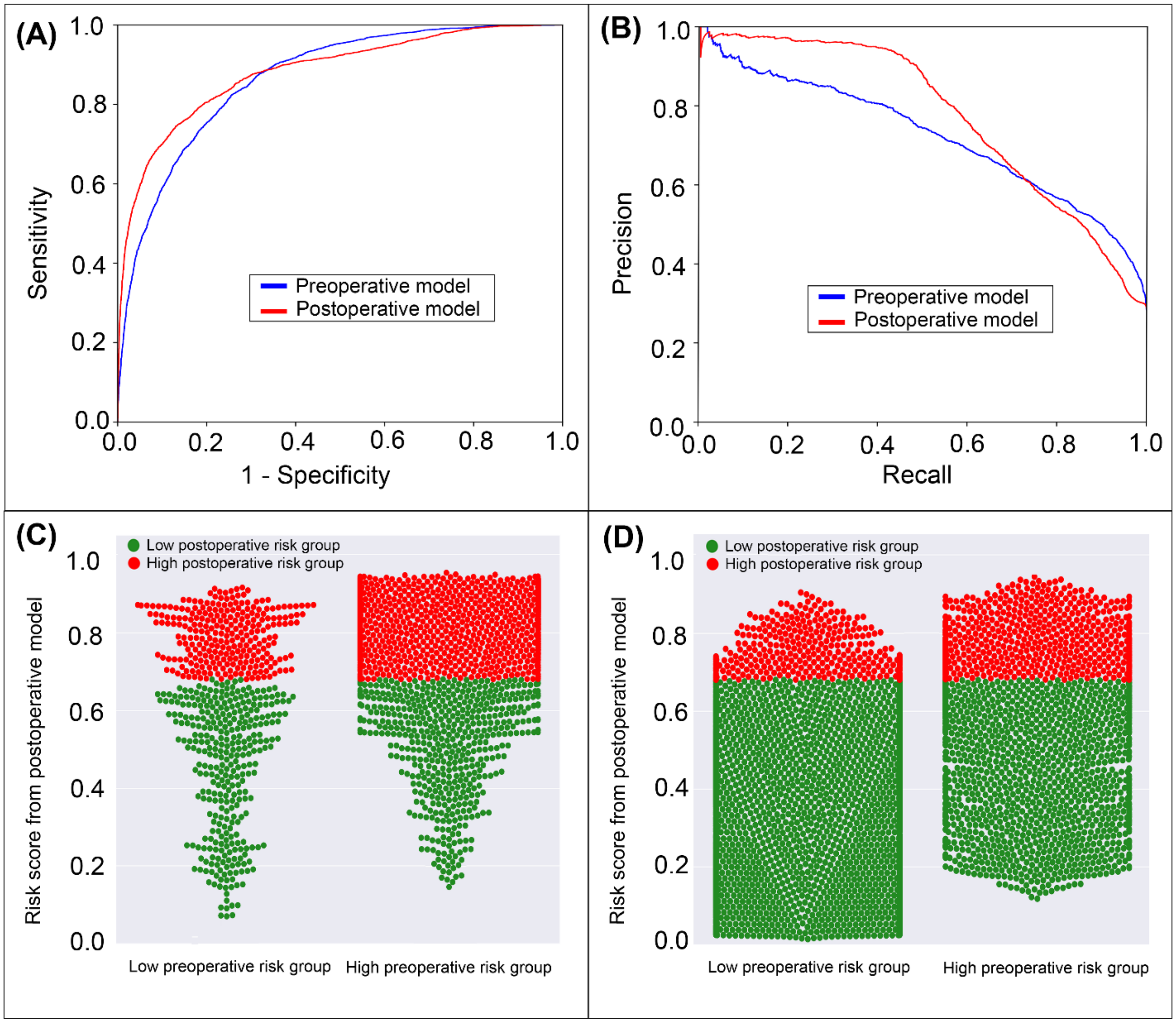

**Figure 4: A model using both preoperative and intraoperative data outperformed a model using preoperative data alone in predicting postoperative mechanical ventilation >48 hours.** A: The postoperative model had greater area under the receiver operating characteristic curve (0.96 vs. 0.89). B: The postoperative model had greater area under the precision-recall curve (0.71 vs. 0.45). The postoperative model reclassified positive cases of prolonged mechanical ventilation duration (C) and negative cases (D). The red dots are patients at high-risk for prolonged mechanical ventilation duration according to the postoperative model, whereas the green dots are patients at low-risk. (C & D) The proposed postoperative model effectively reclassified 10.0% of all cases.

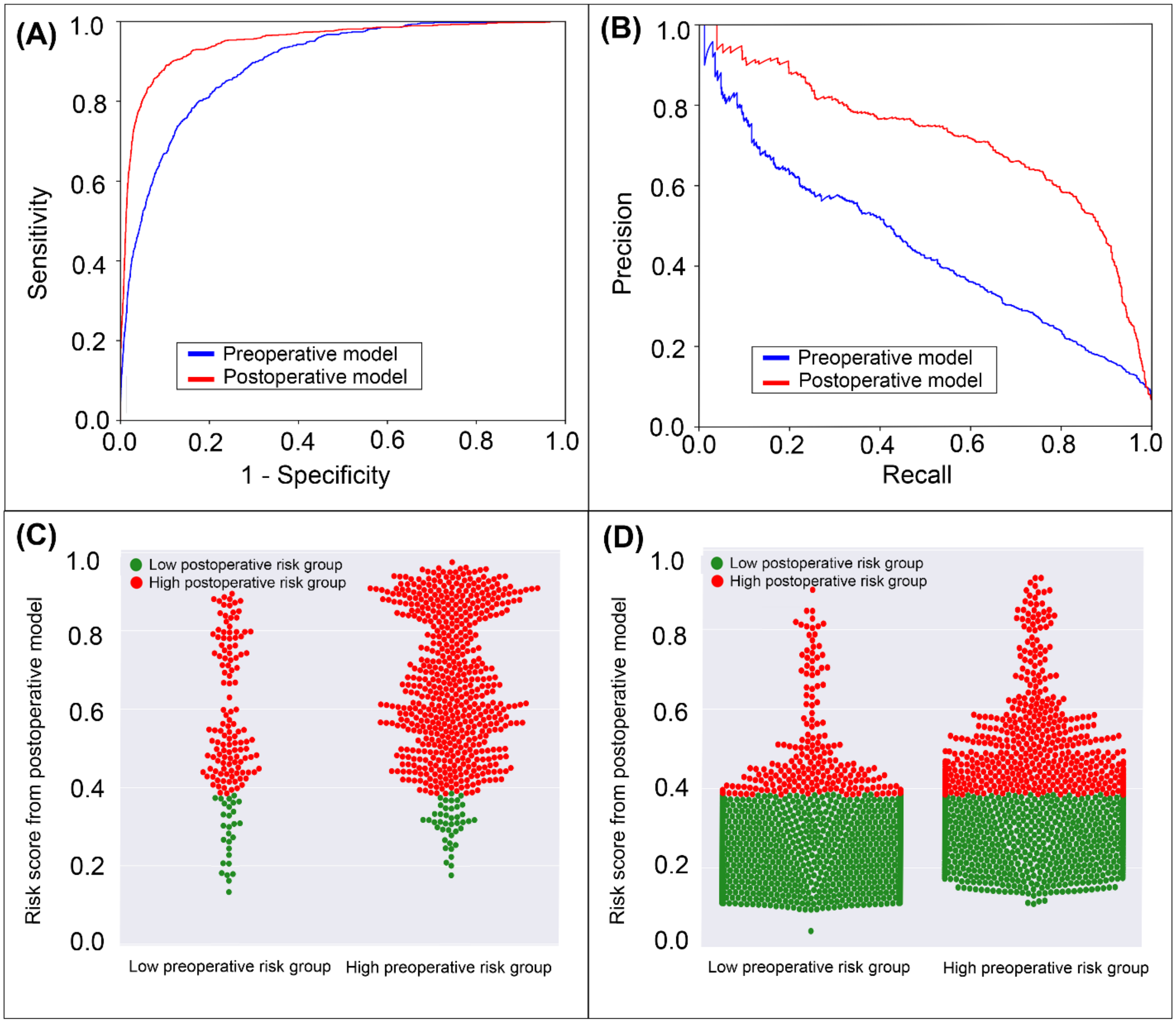

**Figure 5: A model using both preoperative and intraoperative data outperformed a model using preoperative data alone in predicting postoperative neurological complications and delirium.** A: The postoperative model had greater area under the receiver operating characteristic curve (0.89 vs. 0.86). B: The postoperative model had greater area under the precision-recall curve (0.69 vs. 0.64). The postoperative model reclassified positive cases of neurological complication (C) and negative cases (D). The red dots are patients at high-risk for complications according to the postoperative model, whereas the green dots are patients at low-risk. (C & D) The proposed postoperative model effectively reclassified 2.9% of all cases.

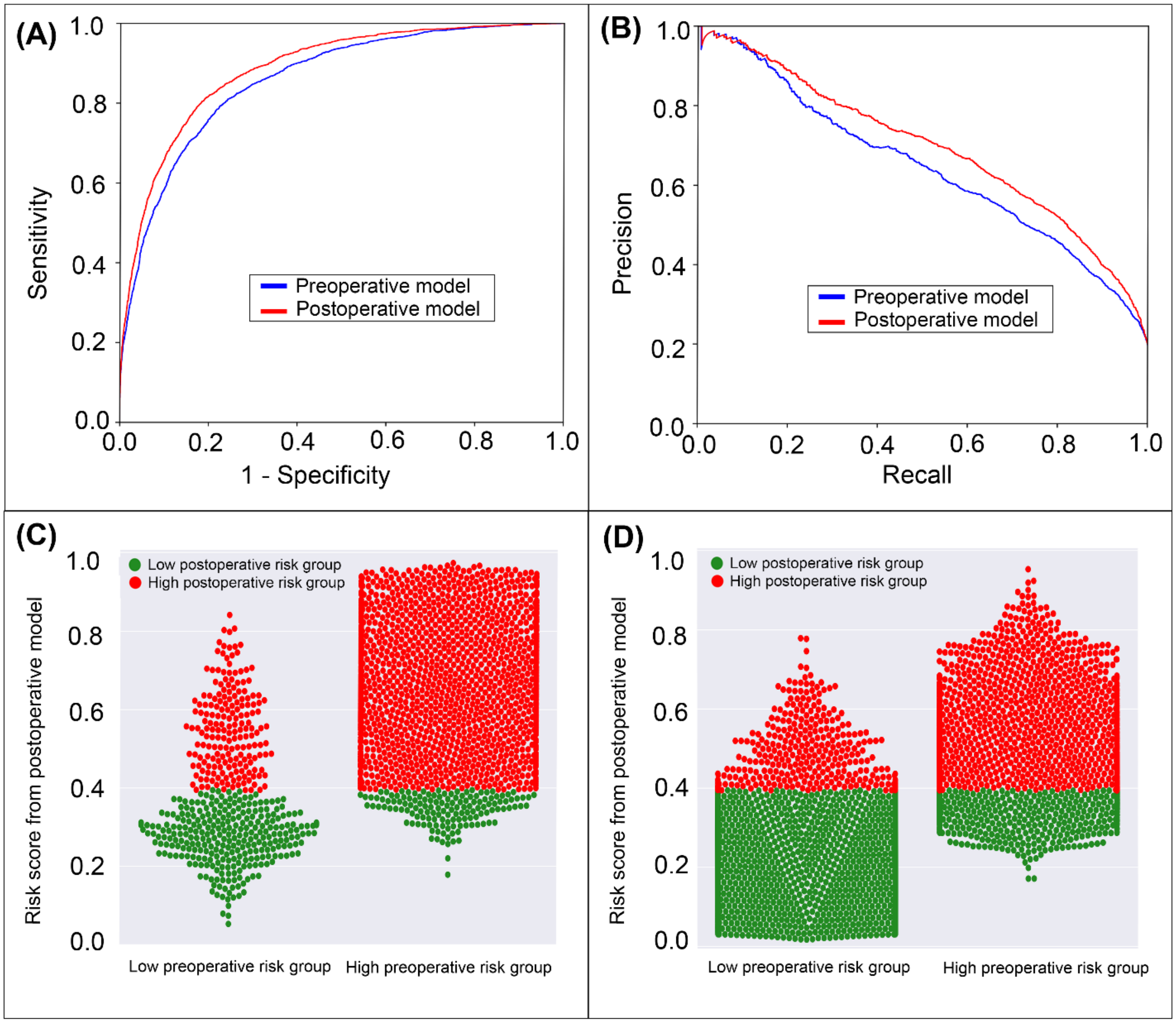

**Figure 6: A model using both preoperative and intraoperative data outperformed a model using preoperative data alone in predicting postoperative hospital mortality.** A: The postoperative model had greater area under the receiver operating characteristic curve (0.93 vs. 0.87). B: The postoperative model had greater area under the precision-recall curve (0.21 vs. 0.15).  The postoperative model reclassified positive cases of hospital mortality (C) and negative cases (D). The red dots are patients at high-risk for mortality according to the postoperative model, whereas the green dots are patients at low-risk. (C & D) The proposed postoperative model effectively reclassified 11.2% of all cases.

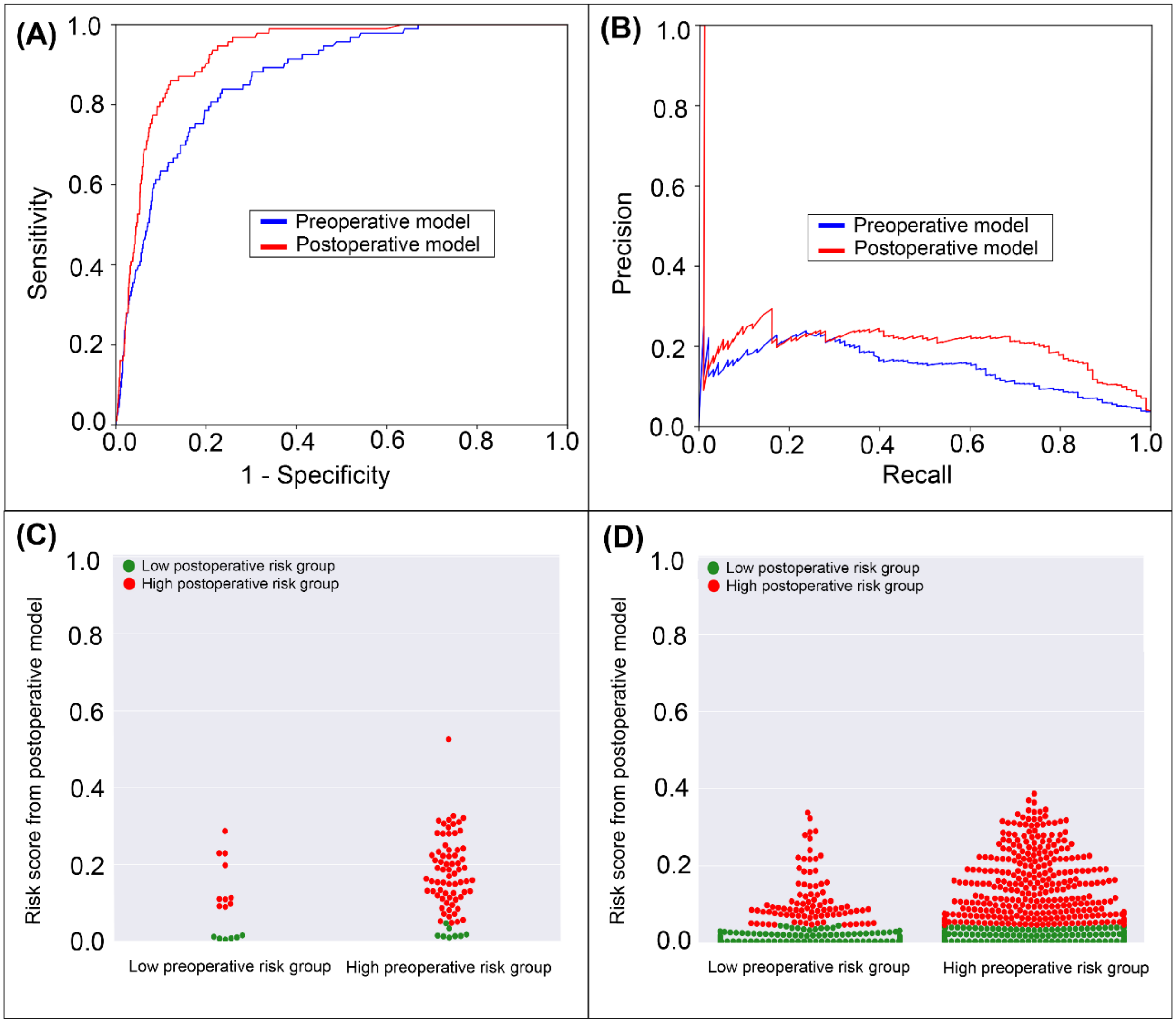

**Table 1: Characteristics of Training and Testing Cohorts.**

| | | **Training** | **Testing** |
|---|---|---|---|
| Date ranges | | June 2014-Feb 2018 (n=40560) | March 2018-Feb 2019 (n=11969) |
| Average age (years) | | 56.5 | 57.5 |
| Ethnicity, n (%) | Not Hispanic | 38116 (93.9) | 11210 (93.6) |
| | Hispanic | 1772 (4.4) | 599 (5) |
| | Missing | 717 (1.8) | 171 (1.4) |
| Race, n (%) | White | 31399 (77.3) | 9376 (78.3) |
| | African American | 6136 (15.1) | 1739 (14.5) |
| | Other | 2483 (6.1) | 702 (5.9) |
| | Missing | 587 (1.5) | 163 (1.4) |
| Gender, n (%) | Male | 20614 (50.8) | 6072 (50.7) |
| | Female | 19991 (49.2) | 5908 (49.3) |
| Primary Insurance, n (%) | Medicare | 18581 (45.8) | 5774 (48.2) |
| | Private | 12463 (30.7) | 3308 (27.6) |
| | Medicaid | 6577 (16.2) | 1928 (16.1) |
| | Uninsured | 2984 (7.4) | 970 (8.1) |
| Outcomes, n (%) | ICU Stay > 48 hours | 10213 (25.2) | 3382 (28.3) |
| | MV Duration > 48 hours | 2372 (5.9) | 767 (6.4) |
| | Neurological Complications and Delirium | 5860 (14.5) | 2364 (19.8) |
| | Hospital Mortality[a] | 192 (2.3) | 93 (2.6) |

[a] Models for hospital mortality were developed using 8,378 surgeries and validated using 35,91surgeries among 11,969 surgeries in the test cohort.

**Table 2: Performance metrics for models predicting postoperative complications and mortality using only preoperative data (preoperative model) and preoperative and intraoperative data (postoperative model).**

| Complication | Model | Sensitivity | Specificity | NPV | PPV | Accuracy | AUROC | AUPRC |
|---|---|---|---|---|---|---|---|---|
| ICU Stay > 48 hours | Preoperative | 0.82 (0.81-0.83) | 0.74 (0.74-0.75) | 0.91 (0.91-0.92) | 0.56 (0.55-0.57) | 0.77 (0.76-0.77) | 0.87 (0.86-0.87) | 0.72 (0.71-0.74) |
| | Postoperative | 0.75 (0.73-0.76) | 0.87 (0.86-0.87) | 0.90 (0.89-0.90) | 0.69 (0.68-0.70) | 0.83 (0.83-0.84) | 0.88 (0.88-0.89) | 0.80 (0.78-0.81) |
| MV Duration > 48 hours | Preoperative | 0.80 (0.78-0.82) | 0.82 (0.82-0.83) | 0.98 (0.98-0.99) | 0.24 (0.22-0.25) | 0.82 (0.81-0.83) | 0.89 (0.87-0.89) | 0.45 (0.42-0.48) |
| | Postoperative | 0.91 (0.89-0.93) | 0.92 (0.92-0.92) | 0.99 (0.99-1.00) | 0.45 (0.41-0.45) | 0.92 (0.91-0.92) | 0.96 (0.95-0.97) | 0.71 (0.68-0.74) |
| Neurological Complications and Delirium | Preoperative | 0.79 (0.77-0.80) | 0.78 (0.77-0.78) | 0.94 (0.93-0.94) | 0.47 (0.45-0.48) | 0.78 (0.77-0.79) | 0.86 (0.85-0.87) | 0.64 (0.63-0.66) |
| | Postoperative | 0.81 (0.80-0.82) | 0.81 (0.80-0.82) | 0.95 (0.94-0.95) | 0.51 (0.49-0.53) | 0.81 (0.79-0.82) | 0.89 (0.88-0.89) | 0.69 (0.67-0.71) |
| Hospital Mortality | Preoperative | 0.83 (0.73-0.87) | 0.76 (0.78-0.80) | 0.99 (0.99-1.00) | 0.09 (0.08-0.11) | 0.77 (0.77-0.80) | 0.87 (0.84-0.90) | 0.15 (0.12-0.20) |
| | Postoperative | 0.85 (0.80-0.91) | 0.88 (0.86-0.88) | 1.00 (0.99-1.00) | 0.16 (0.13-0.18) | 0.88 (0.86-0.88) | 0.93 (0.91-0.95) | 0.21 (0.17-0.27) |

ICU: intensive care unit, MV: mechanical ventilation NPV: negative predictive value, PPV: positive predictive value, AUROC: area under the receiver operating characteristic curve, AUPRC: area under the precision-recall curve.

**Table 3: Net Reclassification Index (NRI) and index improvement analysis for complications and outcomes in study**

| | | | Classification Improvement (%) | | |
|---|---|---|---|---|---|
| **Complication** | **NRI (95% CI)** | **P- value** | **Event** | **Non-Event** | **Overall** |
| ICU stay > 48 hours | 0.05 (0.03-0.06) | <0.001 | -7.9 | 12.6 | 6.8 |
| MV duration > 48 hours | 0.21 (0.16-0.22) | <0.001 | 10.9 | 9.9 | 10.0 |
| Neurological Complications and Delirium | 0.05 (0.03-0.07) | <0.001 | 2.1 | 3.1 | 2.9 |
| Hospital Mortality | 0.14 (0.06-0.21) | 0.024 | 2.2 | 11.5 | 11.2 |

ICU: intensive care unit, MV: mechanical ventilation.

**Supplemental Digital Contents:**

**Supplemental Digital Content 1:** Figure illustrating derivation of the study population.

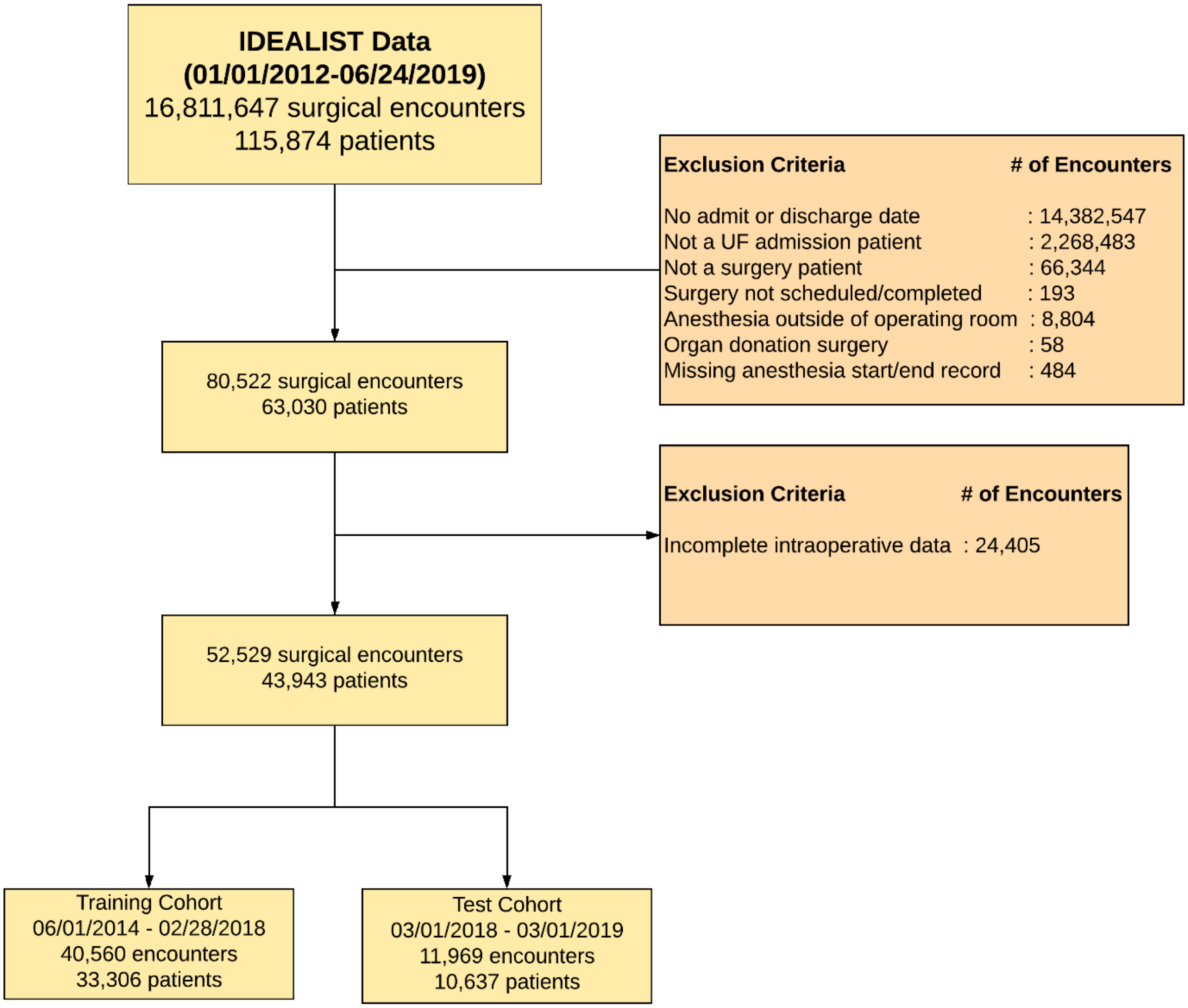

**Supplemental Digital Content 2:** Figure illustrating purpose of data cohorts.

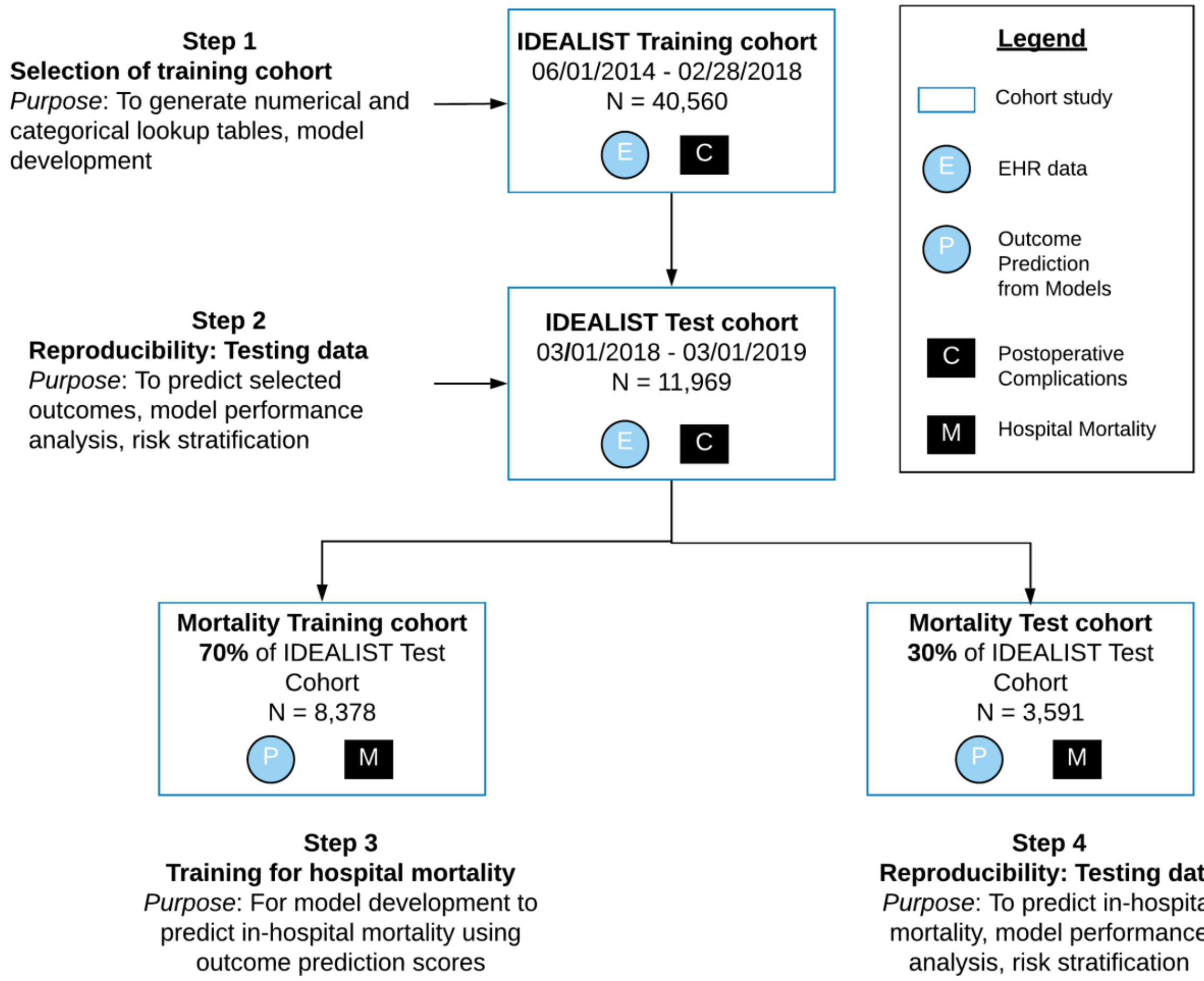

**Supplemental Digital Content 3:** Table listing characteristics of input variables.

| Variable | Type of Variable | Data Source | Number of Categories | Type of Preprocessing |
|---|---|---|---|---|
| **Demographic Variables** | | | | |
| Age (years) | Continuous | Derived | | Imputation of outliers[a] |
| Gender | Binary | Raw | 2 | |
| Race | Nominal | Raw | 5 | Optimization of categorical features[b] |
| Body Mass Index | Continuous | Raw | | Imputation of outliers[a] |
| Marital Status | Nominal | Raw | 3 | Optimization of categorical features[b] |
| Ethnicity | Binary | Raw | 2 | |
| **Socioeconomic Variables** | | | | |
| Primary Insurance | Nominal | Raw | 4 | Optimization of categorical features[b] |
| Residency area characteristics | | | | |
| Zip code | Nominal | Raw | 1,908 | Transformation through link to Census data[c] |
| Rural area | Binary | Derived | 2 | |
| Total Population | Continuous | Derived | | Obtained using residency zip code with linkage to US Census data[c]; Imputation of outliers[a] |
| Median Income | Continuous | Derived | | Obtained using residency zip code with linkage to US Census data[c]; Imputation of outliers[a] |
| Total Proportion of African-Americans | Continuous | Derived | | Obtained using residency zip code with linkage to US Census data[c]; Imputation of outliers[a] |
| Total Proportion of Hispanic | Continuous | Derived | | Obtained using residency zip code with linkage to US Census data[c]; Imputation of outliers[a] |
| Population Proportion Below Poverty | Continuous | Derived | | Obtained using residency zip code with linkage to US Census data[c]; Imputation of outliers[a] |
| Distance from Residency to Hospital (km) | Continuous | Derived | | Obtained using residency zip code with linkage to US Census data[c]; Imputation of outliers[a] |
| **Operative Characteristics** | | | | |
| Day of admission | Nominal | Derived | 7 | Optimization of categorical features[b] |

| Variable | Type of Variable | Data Source | Number of Categories | Type of Preprocessing |
|---|---|---|---|---|
| Month of admission | Nominal | Derived | 12 | Optimization of categorical features[b] |
| Attending Surgeon | Nominal | Raw | 311 | Optimization of categorical features[b] |
| Admission Source | Binary | Raw | 2 | |
| Admission Type (Emergent/Elective) | Binary | Derived | 2 | |
| Admitting Type (Medicine/Surgery) | Binary | Derived | 2 | |
| Admitting Service | Nominal | Derived | 46 | Optimization of categorical features[b] |
| Night Admission | Binary | Derived | 2 | |
| Scheduled Surgery Type | Nominal | Derived | 15 | Optimization of categorical features[b] |
| Scheduled Surgery Room | Nominal | Raw | 41 | Optimization of categorical features[b] |
| Scheduled post operation location | Binary | Derived | 2 | |
| Scheduled room is trauma room | Binary | Derived | 2 | |
| Time of surgery from admission (days) | Continuous | Derived | | Imputation of outliers[a] |
| Scheduled primary surgical procedure | Nominal | Derived | 1555 | Forest tree analysis of ICD9 codes[d] |
| **Comorbidities** | | | | |
| Charlson's Comorbidity Index | Nominal | Derived | 18 | Optimization of categorical features[b] |
| Myocardial Infarction | Binary | Derived | 2 | |
| Congestive Heart Failure | Binary | Derived | 2 | |
| Peripheral Vascular Disease | Binary | Derived | 2 | |
| Cerebrovascular Disease | Binary | Derived | 2 | |
| Chronic Pulmonary Disease | Binary | Derived | 2 | |
| Diabetes | Binary | Derived | 2 | |
| Cancer | Binary | Derived | 2 | |
| Liver Disease | Binary | Derived | 2 | |
| Valvular disease | Binary | Derived | 2 | |
| Coagulopthy | Binary | Derived | 2 | |
| Weight loss | Binary | Derived | 2 | |
| Alcohol or Drug Abuse | Binary | Derived | 2 | |
| Smoking Status | Nominal | Raw | 4 | Optimization of categorical features[b] |
| **Medications History[e]** | | | | |
| Betablockers | Binary | Derived | 2 | |
| Diuretics | Binary | Derived | 2 | |

| Variable | Type of Variable | Data Source | Number of Categories | Type of Preprocessing |
|---|---|---|---|---|
| Statin | Binary | Derived | 2 | |
| Aspirin | Binary | Derived | 2 | |
| Angiotensin-Converting-Enzyme Inhibitors | Binary | Derived | 2 | |
| Pressors or Inotropes | Binary | Derived | 2 | |
| Bicarbonate | Binary | Derived | 2 | |
| Antiemetic | Binary | Derived | 2 | |
| Aminoglycosides | Binary | Derived | 2 | |
| Vancomycin | Binary | Derived | 2 | |
| Nonsteroidal Anti-inflammatory Drug | Binary | Derived | 2 | |
| **Preoperative Laboratory Results** | | | | |
| Urine Protein, mg/dL | Nominal | Derived | 5 | Optimization of categorical features[b] |
| Urine Hemoglobin, mg/dL | Nominal | Derived | 5 | Optimization of categorical features[b] |
| Urine Glucose, mg/dL | Nominal | Derived | 5 | Optimization of categorical features[b] |
| Urine Erythrocytes, mg/dL | Nominal | Derived | 5 | Optimization of categorical features[b] |
| Serum Glucose, mg/dL | Continuous | Raw | | Imputation of outliers[a] |
| Blood Urea Nitrogen test, mg/dL | Continuous | Raw | | Imputation of outliers[a] |
| Serum Creatinine, mg/dL | Continuous | Raw | | Imputation of outliers[a] |
| Serum Calcium, mmol/L | Continuous | Raw | | Imputation of outliers[a] |
| Serum Sodium, mmol/L | Continuous | Raw | | Imputation of outliers[a] |
| Serum Potassium, mmol/L | Continuous | Raw | | Imputation of outliers[a] |
| Serum Chloride, mmol/L | Continuous | Raw | | Imputation of outliers[a] |
| Serum CO2, mmol/L | Continuous | Raw | | Imputation of outliers[a] |
| Serum White Blood Cell, thou/uL | Continuous | Raw | | Imputation of outliers[a] |
| Mean Corpuscular Hemoglobin in Blood, g/dL | Continuous | Raw | | Imputation of outliers[a] |
| Mean Corpuscular Hemoglobin Concentration in Blood, pg | Continuous | Raw | | Imputation of outliers[a] |
| Erythrocyte Distribution Width Count, % | Continuous | Raw | | Imputation of outliers[a] |
| Serum creatinine, mg/dL | Continuous | Raw | | Imputation of outliers[a] |
| Serum Platelet, thou/uL | Continuous | Raw | | Imputation of outliers[a] |
| Serum Hemoglobin, g/dL | Continuous | Raw | | Imputation of outliers[a] |
| Reference Estimated Glomerular Filtration Rate, mL/min/1.73 m² | Continuous | Derived | | Imputation of outliers[a] |

| Variable | Type of Variable | Data Source | Number of Categories | Type of Preprocessing |
|---|---|---|---|---|
| Urea nitrogen-Creatinine ratio in Serum | Continuous | Derived | | Imputation of outliers[a] |
| **Physiologic Intraoperative Time Series** | | | | |
| Systolic blood pressure, mmHg | Continuous | Raw | | Data cleaning[f]; Imputation of outliers[g]; Statistical features extraction[h] |
| Diastolic blood pressure, mmHg | Continuous | Raw | | Data cleaning[f]; Imputation of outliers[g]; Statistical features extraction[h] |
| Minimum alveolar concentration | Continuous | Raw | | Data cleaning[f]; Imputation of outliers[g]; Statistical features extraction[h] |
| Heart rate, bpm | Continuous | Raw | | Data cleaning[f]; Imputation of outliers[g]; Statistical features extraction[h] |
| Temperature ($^{o}$C) | | | | |
| Peripheral capillary oxygen saturation (SPO2) | Continuous | Raw | | Data cleaning[f]; Imputation of outliers[g]; Statistical features extraction[h] |
| End-tidal CO2 (ETCO2) | Continuous | Raw | | Data cleaning[f]; Imputation of outliers[g]; Statistical features extraction[h] |
| Peak Inspiratory Pressure (PIP) | Continuous | Raw | | Data cleaning[f]; Imputation of outliers[g]; Statistical features extraction[h] |
| Positive End-expiratory Pressure (PEEP) | Continuous | Raw | | Data cleaning[f]; Imputation of outliers[g]; Statistical features extraction[h] |
| Respiratory O2 | Continuous | Raw | | Data cleaning[f]; Imputation of outliers[g]; Statistical features extraction[h] |
| Respiratory Rate | Continuous | Raw | | Data cleaning[f]; Imputation of outliers[g]; Statistical features extraction[h] |
| **Laboratory Results from Surgery** | | | | |
| Fraction of inspired oxygen (FIO2) | Continuous | Raw | | Data cleaning[f]; Imputation of outliers[g]; Statistical features extraction[h] |
| **Other Characteristics** | | | | |
| Duration of surgery, min | Continuous | Derived | | |
| Estimated blood loss, mL | Continuous | Raw | | Missing value imputed by 0 |

| Variable | Type of Variable | Data Source | Number of Categories | Type of Preprocessing |
|---|---|---|---|---|
| Urine output, mL | Continuous | Raw | | Missing value imputed by 0 |

A different set of variables was kept in final models (preoperative or intraoperative) from the input set provided in the table.

[a] For continuous variables, observations that fell in the top and bottom 1% of the distribution were considered as outliers and imputed by neighborhood values (i.e., above 99%) are imputed randomly from a uniform distribution defined over [95%, 99.5%] percentiles and below 1% are imputed randomly from another uniform distribution defined over [0.5%, 5%] percentiles.

[b] For categorical variables with more than two levels, levels were transformed to a numeric value as detailed in Methods section.

[c] Using residency zip code, we linked to US Census data to calculate residing neighborhood characteristics and distance from hospital.

[d] Surgical procedure codes were optimized using forest tree analysis of ICD-9-CM codes as detailed in Methods section.

[e] Medications were taken within one year timeframe prior to surgery using RxNorms data grouped into drug classes according to the US, Department of Veterans Affairs National Drug File-Reference Terminology [26].

[f] We used observations for the first surgery, in case multiple surgeries exist. We averaged values if multiple observations exist at a time point.

[g] Values out of the predefined ranges were removed. Additionally, any sudden peak in the time series were converted to median for each encounter.

[h] We extracted several descriptive statistical measures, i.e., mean and standard deviation of time series, minimum and maximum values observed, time/percentage of time a patient spent in a specific range of values for each of the time series.

**Supplemental Digital Content 4:** Table listing additional study population characteristics that are not listed in **Table 1**.

| | Training Cohort | Test Cohort |
|---|---|---|
| **Surgeries** | 40,560 | 11,969 |
| **Male, n (%)** | 20614 (50.8%) | 6072 (50.7%) |
| **Age, mean (SD)** | 56.5 (17.0) | 57.5 (17.3) |
| **Race, n (%)** | | |
| White | 31399 (77.3%) | 9376 (78.3%) |
| African-American | 6136 (15.1%) | 1739 (14.5%) |
| Hispanic | 2483 (6.1%) | 702 (5.9%) |
| Other | 587 (1.5%) | 163 (1.4%) |
| **Primary insurance group, n (%)** | | |
| Medicare | 18581 (45.8%) | 5774 (48.2%) |
| Private | 12463 (30.7%) | 3308 (27.6%) |
| Medicaid | 6577 (16.2%) | 1928 (16.1%) |
| Uninsured | 2984 (7.4%) | 970 (8.1%) |
| **Socio-economic features** | | |
| Neighborhood characteristics | | |
| Rural area, n (%) | 13986 (34.4%) | 4111 (34.3%) |
| Total population, median (25$^{th}$,75th) | 17599 (10884, 27063) | 17599 (10923, 27063) |
| Median income, median (25$^{th}$,75th) | 40528 (35194, 48430) | 40320 (35244, 48245) |
| Total proportion of African-Americans (%), mean (SD) | 0.16 (0.15) | 0.16 (0.15) |
| Total proportion of Hispanic (%), mean (SD) | 0.08 (0.08) | 0.08 (0.08) |
| Population proportion below poverty (%), mean (SD) | 20 (9.5) | 20(2.5) |
| Distance from residency to hospital (km), median (25$^{th}$,75th) | 43.2 (22.1, 81.1) | 43.6 (22.3, 80.7) |
| **Comorbidity features** | | |
| Cancer, n (%) | 11381 (28%) | 3136 (26.2%) |
| Peripheral vascular disease, n (%) | 9207 (22.7%) | 3022 (25.2%) |
| Cerebrovascular disease, n (%) | 7048 (17.4%) | 2221 (18.5%) |
| Myocardial infarction, n (%) | 3187 (7.9%) | 1107 (9.2%) |
| Liver disease, n (%) | 5979 (14.7%) | 1955 (16.3%) |
| Weight Loss, n (%) | 5754 (14.2%) | 2200 (18.4%) |
| Diabetes, n (%) | 10079 (24.8%) | 2917 (24.4%) |
| Alcohol/ Drug abuse, n (%) | 6338 (15.6%) | 1783 (14.9%) |
| Congestive heart failure, n (%) | 7020 (17.3%) | 2343 (19.6%) |
| Valvular Disease, n (%) | 5814 (14.3%) | 2117 (17.7%) |
| Coagulapty, n (%) | 5940 (14.6%) | 1776 (14.8%) |
| Smoking, n (%) | | |

| | Training Cohort | Test Cohort |
|---|---|---|
| Never | 17325 (42.7%) | 5218 (43.6%) |
| Former | 13827 (34.1%) | 4168 (34.8%) |
| Current | 7385 (18.2%) | 2001 (16.7%) |
| Missing | 2068 (5.1%) | 593 (5%) |
| Number of diagnoses, median (25th-75th) | 41 (20, 91) | 48 (25, 110) |
| **Operative features** | | |
| Night admission, n (%) | 18968 (46.7%) | 5718 (47.7%) |
| Admission day (top 3 categories), n (%) | | |
| Monday | 7944 (19.6%) | 2406 (20.1%) |
| Tuesday | 7665 (18.9%) | 2177 (18.2%) |
| Wednesday | 6896 (17%) | 2081 (17.4%) |
| Admission month (top 3 categories), n (%) | | |
| October | 3799 (9.4%) | 1083 (9%) |
| January | 3720 (9.2%) | 1059 (8.8%) |
| August | 3704 (9.1%) | 1049 (8.8%) |
| Number of operating surgeons, n | 283 | 195 |
| Number of procedures per operating surgeon, n (%) | | |
| First rank | 1319 (3.3%) | 316 (2.6%) |
| Second rank | 1129 (2.8%) | 302 (2.5%) |
| Third rank | 1044 (2.6%) | 300 (2.5%) |
| Admission source, n (%) | | |
| Transfer | 6668 (16.4%) | 2046 (17.1%) |
| Emergent at Admission status, n (%) | 15348 (37.8%) | 4855 (40.5%) |
| Admission to surgical service, n (%) | 19264 (47.4%) | 5851 (48.8%) |
| Time of surgery from admission (days), median ($25^{th}$,$75^{th}$) | 3 (2, 25) | 3 (2, 30) |
| **Type of Surgery, n (%)** | | |
| Orthopedic surgery | 9983 (24.6%) | 2895 (24.1%) |
| Neurosurgery | 5625 (13.8%) | 1906 (15.9%) |
| Vascular Surgery | 4043 (10%) | 1190 (9.9%) |
| Thoracic/Cardiovascular surgery | 3214 (7.9%) | 1241 (10.4%) |
| Urologic surgery | 3131 (7.7%) | 704 (5.9%) |
| Trauma- Acute Care surgery | 2953 (7.3%) | 1056 (8.8%) |
| Gastrointestinal surgery | 2521 (6.2%) | 784 (6.5%) |
| Ear, nose, throat surgery | 2446 (6%) | 592 (4.9%) |
| Gynecology obstetrics surgery | 1672 (4.1%) | 320 (2.7%) |
| Pancreas & Biliary, BMSE surgery | 1451 (3.6%) | 337 (2.8%) |
| Transplant surgery | 961 (2.4%) | 227 (1.9%) |
| Plastic surgery | 946 (2.3%) | 216 (1.8%) |
| Burn Surgery | 837 (2.1%) | 185 (1.5%) |

| | Training Cohort | Test Cohort |
|---|---|---|
| Pediatric surgery | 447 (1.1%) | 123 (1%) |
| Other specialty surgeries | 299 (0.7%) | 164 (1.4%) |
| Ophthalmology surgery | 80 (0.2%) | 51 (0.4%) |
| Medicine Gastroenterology | 27 (0.1%) | 4 (0%) |
| **Preoperative and admission day laboratory results** median (25th,75th) | | |
| Hemoglobin within 7 days prior to surgery, g/dl | | |
| Minimum | 12.3 (11.4, 13.3) | 12.3 (11.4, 13.3) |
| Maximum | 13 (12.2, 13.9) | 13 (12.3, 14) |
| Average | 12.7 (11.8, 13.5) | 12.7 (11.9, 13.6) |
| Variance | 0.4 (0.4, 0.4) | 0.4 (0.4, 0.4) |
| Hemoglobin within 8-365 days prior to surgery, g/dl | | |
| Minimum | 11.8 (11.3, 12.2) | 11.8 (11.3, 12.5) |
| Maximum | 13.6 (13.3, 13.8) | 13.6 (13.3, 14) |
| Average | 12.5 (12.2, 12.8) | 12.5 (12.2, 13) |
| Variance | 0.9 (0.9, 0.9) | 0.9 (0.9, 0.9) |
| Glucose in blood within 7 days prior to surgery, mg/dL | | |
| Minimum | 99 (91, 110) | 99 (93, 112) |
| Maximum | 121 (105, 141) | 121 (107, 145) |
| Average | 111 (100, 125) | 111 (102, 126.5) |
| Variance | 269.8 (269.8, 269.8) | 269.8 (269.8, 269.8) |
| Count | 1 (1, 3) | 1 (1, 3) |
| Urea nitrogen in blood within 7 days prior to surgery, mg/dL | | |
| Minimum | 14 (12, 17) | 14 (12, 17) |
| Maximum | 16 (14, 19) | 16 (14, 20) |
| Average | 15 (13, 18) | 15 (13, 18.5) |
| Variance | 4.5 (4.5, 4.5) | 4.5 (4.5, 4.5) |
| Count | 1 (0, 2) | 1 (0, 2) |
| Serum creatinine within 7 days prior to surgery, mg/dL | | |
| Minimum | 0.9 (0.8, 1) | 0.9 (0.7, 0.9) |
| Maximum | 0.9 (0.8, 1.1) | 0.9 (0.8, 1) |
| Average | 0.9 (0.8, 1) | 0.9 (0.8, 1) |
| Variance | 0 (0, 0) | 0 (0, 0) |
| Count | 1 (0, 2) | 1 (0, 3) |
| Serum Calcium within 7 days prior to surgery, mmol/L | | |
| Minimum | 9.1 (8.8, 9.3) | 9.1 (8.8, 9.4) |
| Maximum | 9.3 (9.2, 9.5) | 9.3 (9.2, 9.6) |
| Average | 9.2 (9, 9.4) | 9.2 (9, 9.4) |

| | Training Cohort | Test Cohort |
|---|---|---|
| Variance | 0.1 (0.1, 0.1) | 0.1 (0.1, 0.1) |
| Count | 1 (0, 2) | 1 (0, 2) |
| Serum Sodium ion within 7 days prior to surgery, mmol/L | | |
| Minimum | 138 (137, 139) | 138 (136, 139) |
| Maximum | 140 (139, 141) | 140 (138, 140) |
| Average | 139 (138, 140) | 139 (137, 139) |
| Variance | 2.6 (2.6, 2.6) | 2.6 (2.6, 2.6) |
| Count | 1 (0, 2) | 1 (0, 3) |
| Urea nitrogen-Creatinine ratio within 7 days prior to surgery | | |
| Minimum | 15 (12.9, 17.4) | 15 (13.7, 18.4) |
| Maximum | 17.3 (15, 20) | 17.3 (16, 21.3) |
| Average | 16.2 (14, 18.6) | 16.2 (15, 19.8) |
| Variance | 1.6 (1.6, 1.6) | 1.6 (0, 1.6) |
| Count | 1 (0, 4) | 2 (0, 6) |
| Potassium in serum within 7 days prior to surgery, mmol/L | | |
| Minimum | 3.9 (3.7, 4.1) | 3.9 (3.7, 4) |
| Maximum | 4.2 (4, 4.4) | 4.2 (4, 4.3) |
| Average | 4.1 (3.9, 4.2) | 4.1 (3.9, 4.1) |
| Variance | 0.1 (0.1, 0.1) | 0.1 (0.1, 0.1) |
| Count | 1 (0, 2) | 1 (0, 3) |
| Chloride in Serum within 7 days prior to surgery, mmol/L | | |
| Minimum | 100 (99, 102) | 100 (100, 104) |
| Maximum | 102 (101, 104) | 102 (102, 106) |
| Average | 101 (100, 102.8) | 101 (101, 104.7) |
| Variance | 4 (4, 4) | 4 (4, 4) |
| Count | 1 (0, 2) | 1 (0, 2) |
| Serum CO2 within 7 days prior to surgery, mmol/L | | |
| Minimum | 24 (23, 25) | 24 (23, 26) |
| Maximum | 26 (25, 27) | 26 (25, 27) |
| Average | 25 (24, 26) | 25 (24, 26) |
| Variance | 2.6 (2.6, 2.6) | 2.6 (2.6, 2.6) |
| Count | 1 (0, 2) | 1 (0, 3) |
| White Blood Cell in blood within 7 days prior to surgery, thou/uL | | |
| Minimum | 7.6 (6.6, 8.6) | 7.6 (6.6, 8.5) |
| Maximum | 8.8 (7.5, 10.3) | 8.8 (7.7, 10.2) |
| Average | 8.3 (7.1, 9.4) | 8.3 (7.2, 9.3) |
| Variance | 1.7 (1.7, 1.7) | 1.7 (1.7, 1.7) |
| Count | 1 (0, 2) | 1 (0, 2) |

| | Training Cohort | Test Cohort |
|---|---|---|
| Mean Corpuscular Volume in blood within 7 days prior to surgery, fL | | |
| Minimum | 90.6 (90.6, 90.6) | 90.6 (90.6, 90.6) |
| Maximum | 91.5 (91.5, 91.5) | 91.5 (91.5, 91.5) |
| Average | 91 (91, 91) | 91 (91, 91) |
| Variance | 0.8 (0.8, 0.8) | 0.8 (0.8, 0.8) |
| Count | 0 (0, 1) | 0 (0, 0) |
| Mean Corpuscular Hemoglobin in blood within 7 days prior to surgery, g/dL | | |
| Minimum | 29.8 (29, 30.6) | 29.8 (29.2, 30.8) |
| Maximum | 30.2 (29.4, 31) | 30.2 (29.6, 31.2) |
| Average | 30 (29.2, 30.8) | 30 (29.4, 31) |
| Variance | 0.1 (0.1, 0.1) | 0.1 (0.1, 0.1) |
| Count | 1 (0, 2) | 1 (0, 2) |
| Amount of hemoglobin relative to the size of the cell in blood, g/dL | | |
| Minimum | 32.7 (32.1, 33.2) | 32.9 (32.7, 33.8) |
| Maximum | 33.5 (33, 33.9) | 33.5 (33.5, 34.2) |
| Average | 33.1 (32.6, 33.5) | 33.2 (33.1, 34) |
| Variance | 0.2 (0.2, 0.3) | 0.2 (0.2, 0.2) |
| Count | 2 (0, 4) | 1 (0, 2) |
| Red cell distribution width in Blood within 7 days prior to surgery, % | | |
| Minimum | 14.2 (13.7, 14.7) | 14.2 (13.7, 14.8) |
| Maximum | 14.5 (14, 15.1) | 14.5 (13.9, 15) |
| Average | 14.3 (13.8, 14.9) | 14.3 (13.8, 14.9) |
| Variance | 0.1 (0.1, 0.1) | 0.1 (0.1, 0.1) |
| Count | 1 (0, 2) | 1 (0, 2) |
| Platelet in blood, within 7 days prior to surgery thou/uL | | |
| Minimum | 219 (192, 248) | 219 (194, 250) |
| Maximum | 239 (211, 269) | 239 (215, 273) |
| Average | 228 (202, 258) | 228 (205, 259.5) |
| Variance | 406.9 (406.9, 406.9) | 406.9 (406.9, 406.9) |
| Count | 1 (0, 2) | 1 (0, 2) |
| Mean platelet volume in blood within 7 days prior to surgery, fL | | |
| Minimum | 7.8 (7.8, 7.8) | 7.8 (7.8, 7.8) |
| Maximum | 8.3 (8.3, 8.3) | 8.3 (8.3, 8.3) |
| Average | 8 (8, 8) | 8 (8, 8) |
| Variance | 0.2 (0.2, 0.2) | 0.2 (0.2, 0.2) |
| Count | 0 (0, 1) | 0 (0, 0) |
| Reference estimated glomerular filtration rate | 92.9 (83, 102.7) | 92.9 (82.5, 103.3) |

| | Training Cohort | Test Cohort |
|---|---|---|
| [a]Automated urinalysis, urine protein within 365 days prior to surgery (mg/dL), n (%) | | |
| Missing | 21410 (52.7%) | 6631 (55.3%) |
| Negative | 12424 (30.6%) | 3503 (29.2%) |
| Small (<30) | 1189 (2.9%) | 120 (1%) |
| Moderate (300) | 4423 (10.9%) | 1352 (11.3%) |
| Large (>=300) | 1194 (2.9%) | 389 (3.2%) |
| [a]Automated urinalysis, urine glucose within 7 days prior to surgery (mg/dL), n (%) | | |
| Missing | 30673 (75.5%) | 9299 (77.2%) |
| Negative | 8740 (21.5%) | 2347 (19.6%) |
| Small (<499) | 661 (1.6%) | 200 (1.7%) |
| Moderate (1000) | 317 (0.8%) | 138 (1.2%) |
| Large (>1000) | 249 (0.6%) | 11 (0.1%) |
| [a]Automated urinalysis, urine glucose within 8 to 365 days prior to surgery (mg/dL), n (%) | | |
| Missing | 28151 (69.3%) | 8492 (70.8%) |
| Negative | 11446 (28.2%) | 3195 (26.6%) |
| Small (<500) | 440 (1.1%) | 137 (1.1%) |
| Moderate (<1000) | 269 (0.7%) | 140 (1.2%) |
| Large (>1000) | 334 (0.8%) | 31 (0.3%) |
| [b]Automated urinalysis, unire hemoglobin within 7 days prior to surgery (mg/dL), n (%) | | |
| Missing | 34190 (84.2%) | 11970 (99.9%) |
| Negative | 4030 (9.9%) | 7 (0.1%) |
| Small | 1266 (3.1%) | 0 (0%) |
| Moderate | 620 (1.5%) | 1 (0%) |
| Large | 499 (1.2%) | 2 (0%) |
| [b]Automated urinalysis, urine hemoglobin within 8 to 365 days prior to surgery (mg/dL), n (%) | | |
| Missing | 32585 (80.3%) | 11812 (98.6%) |
| Negative | 5860 (14.4%) | 134 (1.1%) |
| Small | 956 (2.4%) | 17 (0.1%) |
| Moderate | 548 (1.4%) | 7 (0.1%) |
| Large | 656 (1.6%) | 10 (0.1%) |
| [a]Automated urinalysis, urine erythrocytes within 365 days prior to surgery (mg/dL), n (%) | | |
| Missing | 24724 (60.8%) | 7163 (59.7%) |
| Negative (<=4) | 12657 (31.1%) | 4099 (34.2%) |
| Small (>4) | 1423 (3.5%) | 175 (1.5%) |
| Moderate (>30) | 411 (1%) | 192 (1.6%) |

| | Training Cohort | Test Cohort |
|---|---|---|
| Large (>=50) | 1425 (3.5%) | 366 (3.1%) |
| Number of complete blood count tests, n (%) | 29021(71.5%) | 8436 (70.4%) |
| **Medication history (1 year prior to Surgery)** | | |
| Medication groups, n (%) | | |
| Beta blockers | 6994 (17.2%) | 2153 (18%) |
| Diuretics | 4602 (11.3%) | 1323 (11%) |
| Statins | 3676 (9.1%) | 1259 (10.5%) |
| Aspirin | 5708 (14.1%) | 1807 (15.1%) |
| Angiotensin-converting-enzyme inhibitors | 4139 (10.2%) | 1204 (10.1%) |
| Vasopressors and inotropes | 8427 (20.8%) | 2799 (23.4%) |
| Bicarbonate | 4582 (11.3%) | 1420 (11.9%) |
| Anti-emetics | 11788 (29%) | 3694 (30.8%) |
| Aminoglycosides | 1371 (3.4%) | 463 (3.9%) |
| **Intraoperative Variables[c]** | | |
| Diastolic Blood Pressure, mm Hg | | |
| Minimum, mean (SD) | 40.54 (12.99) | 41.44 (14.25) |
| Maximum, mean (SD) | 98.91 (25.69) | 101.43 (26.18) |
| Average, mean(SD) | 63.57 (9.69) | 65 (10.15) |
| Long Term Variability, mean (SD) | 119.45 (119.92) | 120.78 (105.91) |
| Short Term Variability, mean (SD) | 34.85 (53.71) | 33.58 (54.05) |
| Systolic Blood Pressure, mm Hg | | |
| Minimum, mean (SD) | 76.43 (20.97) | 76.52 (22.41) |
| Maximum, mean (SD) | 165.05 (32.27) | 167.86 (33.11) |
| Average, mean(SD) | 115.09 (14.66) | 116.36 (14.83) |
| Long Term Variability, mean (SD) | 315.68 (248.99) | 324.53 (247.35) |
| Short Term Variability, mean (SD) | 45.97 (67.87) | 45.63 (65.34) |
| Heart Rate, bpm | | |
| Minimum, mean (SD) | 60 (13.41) | 59.99 (13.87) |
| Maximum, mean (SD) | 109.32 (29.49) | 107.99 (27.88) |
| Average, mean(SD) | 77.73 (13.35) | 77.87 (13.5) |
| Long Term Variability, mean (SD) | 108.93 (243.74) | 96.84 (160.45) |
| Short Term Variability, mean (SD) | 8.46 (16.55) | 8.09 (15) |
| Respiratory Rate | | |
| Minimum, mean (SD) | 2.6 (2.34) | 3.11 (3.77) |
| Maximum, mean (SD) | 26.03 (8.59) | 25.91 (8.72) |
| Average, mean(SD) | 11.49 (2.86) | 12.1 (3.09) |
| Long Term Variability, mean (SD) | 13.32 (14.56) | 13.68 (16.34) |
| Short Term Variability, mean (SD) | 1.37 (3.6) | 1.43 (3.67) |
| Peripheral capillary oxygen saturation (SpO2) | | |
| Minimum, mean (SD) | 88.18 (9.37) | 87.78 (9.88) |

| | Training Cohort | Test Cohort |
|---|---|---|
| Maximum, mean (SD) | 99.88 (0.65) | 99.87 (0.7) |
| Average, mean(SD) | 98.12 (1.81) | 98.05 (1.86) |
| Long Term Variability, mean (SD) | 4.84 (13.78) | 5.46 (24.26) |
| Short Term Variability, mean (SD) | 0.15 (1.1) | 0.16 (1.42) |
| End-tidal CO2 (ETCO2) | | |
| Minimum, mean (SD) | 15.71 (5.85) | 17.52 (7.62) |
| Maximum, mean (SD) | 46.72 (8.25) | 45.28 (8.93) |
| Average, mean(SD) | 34.15 (4.67) | 34.12 (5.24) |
| Long Term Variability, mean (SD) | 26.06 (25.99) | 26 (25.84) |
| Short Term Variability, mean (SD) | 1.26 (7.64) | 1.18 (2.54) |
| Respiratory O2 | | |
| Minimum, mean (SD) | 2.74 (1.79) | 2.83 (1.67) |
| Maximum, mean (SD) | 2.79 (1.84) | 2.89 (1.77) |
| Average, mean(SD) | 2.76 (1.8) | 2.86 (1.7) |
| Long Term Variability, mean (SD) | 0.72 (4.91) | 0.75 (5.74) |
| Short Term Variability, mean (SD) | 0 (0) | 0 (0) |
| Fraction of inspired oxygen (FiO2) | | |
| Minimum, mean (SD) | 30.33 (9.09) | 25.23 (7.58) |
| Maximum, mean (SD) | 39.8 (10.16) | 35.83 (13.09) |
| Average, mean(SD) | 37.76 (5.77) | 31.82 (9.13) |
| Long Term Variability, mean (SD) | 5.85 (17.2) | 8.06 (33.41) |
| Short Term Variability, mean (SD) | 0.45 (9.12) | 0.4 (4.64) |
| Positive end-expiratory pressure (PEEP) | | |
| Minimum, mean (SD) | 6.14 (2.22) | 6.2 (2.07) |
| Maximum, mean (SD) | 6.19 (2.24) | 6.24 (2.1) |
| Average, mean(SD) | 6.16 (2.21) | 6.22 (2.07) |
| Long Term Variability, mean (SD) | 2.2 (4.11) | 2.86 (5.69) |
| Short Term Variability, mean (SD) | 0 (0) | 0 (0) |
| Peak Inspiratory Pressure (PIP) | | |
| Minimum, mean (SD) | 0.08 (0.89) | 0.07 (0.83) |
| Maximum, mean (SD) | 25.42 (10.04) | 24.93 (10.08) |
| Average, mean(SD) | 15.42 (7.08) | 14.73 (6.84) |
| Long Term Variability, mean (SD) | 47.01 (36.57) | 45.8 (36.73) |
| Short Term Variability, mean (SD) | 2.89 (4.24) | 2.68 (4.16) |
| Minimum alveolar concentration (MAC) | | |
| Minimum, mean (SD) | 0.07 (0.07) | 0.04 (0.07) |
| Maximum, mean (SD) | 1.06 (0.38) | 1.02 (0.43) |
| Average, mean(SD) | 0.61 (0.22) | 0.55 (0.26) |
| Long Term Variability, mean (SD) | 0.05 (0.04) | 0.06 (0.06) |
| Short Term Variability, mean (SD) | 0 (0.01) | 0 (0.01) |
| Temperature, $^0$C | | |
| Minimum, mean (SD) | 35.49 (2.62) | 35.62 (2.78) |

| | Training Cohort | Test Cohort |
|---|---|---|
| Maximum, mean (SD) | 37.59 (0.69) | 37.57 (0.7) |
| Average, mean(SD) | 36.96 (0.82) | 36.98 (0.86) |
| Variance, mean (SD) | 0.47 (2.08) | 0.68 (2.52) |

[a] Result of both numeric data and text extraction
[b] Result of text extraction; no numerical extraction was performed
[c] Missing values were imputed with median values

**Supplemental Digital Content 5:** Table listing allowed laboratory and time-series variable ranges.

| **Laboratory Test Variables** | **Ranges** |
|---|---|
| Blood sugar test (Glucose), mg/dL | 25-1400 |
| Blood urea nitrogen test, mg/dL | 1-200 |
| Serum Creatinine, mg/dL | 0.1-20 |
| Serum Calcium, mmol/L | 1.0-30 |
| Serum Sodium, mmol/L | 80-190 |
| Serum Potassium, mmol/L | 0-30 |
| Serum Chlorine, mmol/L | 36-150 |
| Serum O2 Saturation, % | 0-100 |
| Serum CO2, mmol/L | 1.0-50 |
| Serum white blood cell, thou/uL | 0.1-240 |
| Erythrocyte Mean Corpuscular Hemoglobin in Blood, pg | 10.0-50 |
| Erythrocyte distribution width count, % | 2.0-40 |
| Serum Platelet, thou/uL | 2.0-1900 |
| Serum Hemoglobin, g/dL | 3.0-23.0 |
| Reference estimated glomerular filtration rate, mL/min/1.73 m² | 0-200 |
| Urea nitrogen-Creatinine ratio in Serum | 0.2-100 |
| Blood bicarbonate, mmol/dL | 3-62 |
| Anion Gap in Serum, mmol/L | 1 to 40 |
| Mean Platelet Valome, fL | 1 to 30 |
| Erythrocyte mean corpuscular volume, fL | 20-150 |
| Erythrocyte mean corpuscular hemoglobin concentration, g/dL | 1-200 |
| Erythrocyte sedimentation rate, mm/h | 1-140 |
| **IntraOperative Time-Series variables** | |
| End-tidal CO2 (ETCO2) | 10-200 |
| Fraction of inspired oxygen (FiO2) | 21-200 |
| Respiratory O2 | 0-20 |
| Positive end-expiratory pressure (PEEP) | 0-30 |
| Peak Inspiratory Pressure (PIP) | 0-40 |
| Respiratory Rate | 0-60 |
| Peripheral capillary oxygen saturation (SpO2) | 0-100 |
| Temperature, $^{0}$C | 24-45 |
| Heart Rate, bpm | 0-300 |
| Systolic Blood Pressure, mm Hg | 20-300 |
| Diastolic Blood Pressure, mm Hg | 5-225 |
| Minimum alveolar concentration (MAC) | 0-2 |

**Supplemental Digital Content 6:** Table listing ICD codes used to identify neurological complications and delirium.

| **ICD9 codes** | **ICD10 codes by Conversion** | **ICD10 codes by clinician** |
|---|---|---|
| | | |
| 331.3, 331.4 Hydrocephalus | G91.0 Communicating hydrocephalus<br>G91.1 Obstructive hydrocephalus | G81.0 Flaccid hemiplegia |
| 342.0 - 342.9 Hemiplegia or Hemiparesis | | 81.1 Spastic hemiplegia |
| 342.0 Flaccid hemiplegia | G81.00 Flaccid hemiplegia affecting unspecified side<br>G81.01 Flaccid hemiplegia affecting right dominant side<br>G81.02 Flaccid hemiplegia affecting left dominant side<br>G81.03 Flaccid hemiplegia affecting right nondominant side<br>G81.04 Flaccid hemiplegia affecting left nondominant side | G81.9 Hemiplegia, unspecified |
| 342.1 Spastic hemiplegia | G81.10 Spastic hemiplegia affecting unspecified side<br>G81.11 Spastic hemiplegia affecting right dominant side<br>G81.12 Spastic hemiplegia affecting left dominant side<br>G81.13 Spastic hemiplegia affecting right nondominant side<br>G81.14 Spastic hemiplegia affecting left nondominant side | G82.2 Paraplegia |
| 342.8 Other specified hemiplegia | G81.90 Hemiplegia, unspecified affecting unspecified side<br>G81.91 Hemiplegia, unspecified affecting right dominant side<br>G81.92 Hemiplegia, unspecified affecting left dominant side<br>G81.93 Hemiplegia, unspecified affecting right nondominant side | G82.5 Quadriplegia |

| ICD9 codes | ICD10 codes by Conversion | ICD10 codes by clinician |
|---|---|---|
| | G81.94 Hemiplegia, unspecified affecting left nondominant side | |
| 342.9 Hemiplegia, unspecified | G81.90 Hemiplegia, unspecified affecting unspecified side<br>G81.91 Hemiplegia, unspecified affecting right dominant side<br>G81.92 Hemiplegia, unspecified affecting left dominant side<br>G81.93 Hemiplegia, unspecified affecting right nondominant side<br>G81.94 Hemiplegia, unspecified affecting left nondominant side | G83 Other paralytic syndromes |
| 348.1 Anoxic brain damage: | G93.1 Anoxic brain damage, not elsewhere classified | G91 Hydrocephalus |
| 434.0 - 434.9 (434.0, 434.1, 434.9) Occlusion of a cerebral artery, with or without infarction | I63.40 Cerebral infarction due to embolism of unspecified cerebral<br>I63.30 Cerebral infarction due to thrombosis of unspecified cerebra<br>I63.50 Cerebral infarction due to unspecified occlusion or stenosis<br>I66.19 Occlusion and stenosis of unspecified anterior cerebral arte<br>I66.9 Occlusion and stenosis of unspecified cerebral artery<br>I66.09 Occlusion and stenosis of unspecified middle cerebral artery<br>I66.29 Occlusion and stenosis of unspecified posterior cerebral art | G93.1 Anoxic brain damage, not elsewhere classified |
| 784.3 Aphasia | R47.01 Aphasia | G94 Other disorders of brain in diseases classified elsewhere |
| 997.00 - 997.09 Postoperative neurological complications, including infarction or hemorrhage | | G95 Other and unspecified diseases of spinal cord |

| ICD9 codes | ICD10 codes by Conversion | ICD10 codes by clinician |
|---|---|---|
| 997.0 Nervous system complications | | G97.0 Cerebrospinal fluid leak from spinal puncture |
| 997.00 Nervous system complication, unspecified | G97.81 Other intraoperative complications of nervous system | G97.1 Other reaction to spinal and lumbar puncture |
| 997.01 Central nervous system complication: | G97.81 Other intraoperative complications of nervous system<br>G97.82 Other postprocedural complications and disorders of nervous | G97.2 Intracranial hypotension following ventricular shunting |
| 997.02 Iatrogenic cerebrovascular infarction or hemorrhage Postoperative stroke | I97.811 Intraoperative cerebrovascular infarction during other surge<br>I97.821 Postprocedural cerebrovascular infarction during other surge | G97.8 Other intraoperative and postprocedural complications and disorders of nervous system |
| 997.09 Other nervous system complications | G03.8 Meningitis due to other specified causes<br>G97.0 Cerebrospinal fluid leak from spinal puncture<br>G97.81 Other intraoperative complications of nervous system<br>G97.82 Other postprocedural complications and disorders of nervous | R47.0 Dysphasia and aphasia |
| | | I60 Nontraumatic subarachnoid hemorrhage |
| | | I61 Nontraumatic intracerebral hemorrhage |
| | | I62 Other and unspecified nontraumatic intracranial hemorrhage |
| | | I63 Cerebral infarction |
| | | I97.81 Intraoperative cerebrovascular infarction |

| ICD9 codes | ICD10 codes by Conversion | ICD10 codes by clinician |
|---|---|---|
| | | I97.82 Postprocedural cerebrovascular infarction |

Delirium was coded by CAM scores.